\documentclass[aps,pra,showpacs,superscriptaddress,preprint,nofootinbib]{revtex4-1}

\usepackage[utf8]{inputenc}
\usepackage{amsmath}
\usepackage{amssymb}
\usepackage{color}
\usepackage{verbatim}
\usepackage{array}
\usepackage{graphicx}
\usepackage{epsfig}
\usepackage{bm}
\usepackage[ragged]{sidecap}
\usepackage{dsfont}
\usepackage{ulem}
\usepackage{mathrsfs}

\usepackage{caption}
\usepackage{subcaption}
\usepackage{gensymb}
\usepackage{siunitx}
\usepackage{url}
\usepackage{hyperref}
\usepackage{dcolumn}
\usepackage{soul}
\newcolumntype{d}{D{.}{.}{-1}}
\newcolumntype{f}[1]{D{.}{.}{#1}}

\newcommand{\mm}{\mathrm{m}}

\newcommand{\etal}{\textit{et al. }}

\newcommand{\ie}{\textit{i.e.}, } 
\newcommand{\eg}{\textit{e.g.}, }

\begin{document}


\title{High-precision measurements of $n=2\to n=1$ transition energies and level widths in He- and Be-like Argon Ions}

%
%
%


%
%

\author{J.\ Machado} 
\affiliation{Laborat\'orio de Instrumenta{\c c}\~ao, Engenharia Biom\'edica e F\'isica da Radia{\c c}\~ao (LIBPhys-UNL), Departamento de F\'isica, Faculdade de Ci\^encias e Tecnologia, \\
 FCT, Universidade Nova de Lisboa, 2829-516 Caparica, Portugal}
\affiliation{Laboratoire Kastler Brossel, Sorbonne Universit\'e, CNRS, ENS-PSL Research University, Coll\`ege de France, Case\ 74;\ 4, place Jussieu, F-75005 Paris, France}

\author{C.\ I.\ Szabo}  
\affiliation{National Institute of Standards and Technology, Gaithersburg, MD20899, USA}
\affiliation{Theiss Research, La Jolla, CA 92037, USA}
\author{J.\ P.\ Santos}  
\affiliation{Laborat\'orio de Instrumenta{\c c}\~ao, Engenharia Biom\'edica e F\'isica da Radia{\c c}\~ao (LIBPhys-UNL), Departamento de F\'isica, Faculdade de Ci\^encias e Tecnologia, \\
 FCT, Universidade Nova de Lisboa, 2829-516 Caparica, Portugal}
\email{jps@fct.unl.pt} 
\author{P.\ Amaro} 
\affiliation{Laborat\'orio de Instrumenta{\c c}\~ao, Engenharia Biom\'edica e F\'isica da Radia{\c c}\~ao (LIBPhys-UNL), Departamento de F\'isica, Faculdade de Ci\^encias e Tecnologia, \\
 FCT, Universidade Nova de Lisboa, 2829-516 Caparica, Portugal}
\author{M.\ Guerra} 
\affiliation{Laborat\'orio de Instrumenta{\c c}\~ao, Engenharia Biom\'edica e F\'isica da Radia{\c c}\~ao (LIBPhys-UNL), Departamento de F\'isica, Faculdade de Ci\^encias e Tecnologia, \\
 FCT, Universidade Nova de Lisboa, 2829-516 Caparica, Portugal}

\author{ A.\ Gumberidze}
\affiliation{ExtreMe Matter Institute EMMI and Research Division, GSI Helmholtzzentrum für Schwerionenforschung, D-64291 Darmstadt, Germany}

\author{ Guojie Bian}  
\affiliation{Laboratoire Kastler Brossel, Sorbonne Universit\'e, CNRS, ENS-PSL Research University, Coll\`ege de France, Case\ 74;\ 4, place Jussieu, F-75005 Paris, France}
\affiliation{Institute of Atomic and Molecular Physics, Sichuan University, Chengdu 610065, P.R. China}
\author{ J.\ M.\ Isac}  
\affiliation{Laboratoire Kastler Brossel, Sorbonne Universit\'e, CNRS, ENS-PSL Research University, Coll\`ege de France, Case\ 74;\ 4, place Jussieu, F-75005 Paris, France}
\author{ P.\ Indelicato}  
\affiliation{Laboratoire Kastler Brossel, Sorbonne Universit\'e, CNRS, ENS-PSL Research University, Coll\`ege de France, Case\ 74;\ 4, place Jussieu, F-75005 Paris, France}
\email{paul.indelicato@lkb.upmc.fr}
%

\date{\today}


\begin{abstract}

We  performed a reference-free measurement of the transition energies of the $1s 2p\,^1P_1\to 1s^2 \,^1S_0$ line in He-like argon,  and of the $1s 2s^2 2p\,^1P_1\to 1s^2 2s^2\,^1S_0$ line in  Be-like argon ions. The highly-charged ions were produced in the plasma of an Electron-Cyclotron Resonance Ion Source. Both energy measurements were performed with an accuracy better than 3 parts in $10^6$, using a double flat-crystal spectrometer, without reference to any theoretical or experimental energy.  The  $1s 2s^2 2p\,^1P_1\to 1s^2 2s^2\,^1S_0$ transition measurement is the first reference-free measurement for this core-excited transition. The   $1s 2p\,^1P_1\to 1s^2 \,^1S_0$  transition measurement confirms recent measurement performed at the Heidelberg Electron-Beam Ion Trap (EBIT). The width measurement in the He-like transition provides test of a purely radiative decay calculation. In the case of the Be-like argon transition, the width results from the sum of a radiative channel and  three main Auger channels. We also performed Multiconfiguration Dirac-Fock (MCDF) calculations of transition energies and rates and have done an extensive comparison with theory and other experimental data. For both measurements reported here, we find agreement with the most recent theoretical calculations within the combined theoretical and experimental uncertainties.

\end{abstract}

%
%
\maketitle
%
%

%

\section{Introduction}
\label{sec:intro}

Bound-states quantum electrodynamics (BSQED) and the relativistic many-body problem have been undergoing  important progress in the past few years. Yet there are several issues that require increasing the number of high-precision tests. High-precision measurements of transition energies on medium to high-$Z$ elements \cite{bbkc2007,asgl2012,ckgh2012,rbes2013,cpgh2014,pckg2014,kmmu2014,bab2015,esbr2015}, Land\'e $g$-Factors \cite{swsz2011,wskg2013,skzw2014,kskw2015,uadg2015,mips2016,ybht2016} and hyperfine structure\cite{bosc1998,sbde1998,bai2000,buwl2001,yas2001,sayz2001,vgas2012,agvs2012,nljg2013,ljga2014,btbc2014,vadg2015,uabd2017}, just to name a few, are needed either to improve our understanding or to provide tests of higher-order QED-corrections, the calculations of which are very demanding. 

Recent measurement of the proton size in muonic hydrogen~\cite{pana2010,ansa2013} and of the deuteron in muonic deuterium~\cite{pnfa2016}, which disagree by 7 and 3.5 standard deviations respectively from measurements in their electronic counterparts triggered experimental and theoretical research regarding not only the specific issue of the proton and deuteron size, but also the possible anomalies in  BSQED.  A discrepancy  of this magnitude corresponds to a difference in the muonic hydrogen energy of \SI{0.42}{\meV},  which is far outside the calculations uncertainty of about \SI{\pm 0.01}{\meV} and is much larger than what can be expected from any omitted QED contribution. Another large discrepancy of 7 standard deviations between theory and experiment has also been observed recently in a specific difference  between the hyperfine structures of hydrogenlike and lithiumlike bismuth measured at the Experimental Storage Ring (ESR) at GSI in Darmstadt \cite{uabd2017}, designed to eliminate the effect of the nuclear magnetization distribution (the Bohr-Weisskopf correction) \cite{sayz2001}.
 
Medium and high-$Z$ few-electron ions with a $K$ hole are the object of the present work. They have been studied first in laser-produced plasmas \cite{abzp1974} and  beam-foil spectroscopy (see, \eg \cite{dam1979,bmic1983}), low-inductance vacuum spark \cite{aamp1988}, or by using the interaction of fast ion beams with gas targets in heavy-ion accelerators. Ion storage rings have also been used (see, \eg \cite{smbb1994,smbd2000,gsbb2005}). The limitation in precision of those measurements is mostly due to the large Doppler effect, which affects  energy measurements, and the Doppler broadening, which affects any possible width measurement.  

Recoil ion spectroscopy \cite{dbf1984}, which has also been used, is not affected by the Doppler effect, and provides an interesting check. Plasma machines, such as tokamaks, have also provided spectra \cite{bhzv1985,hbhv1987}, leading to relative measurements, without Doppler shift, usually using  He-like lines as a reference. Solar measurements \cite{neu1971} have also been reported. 

Accurate transition energy measurements  in medium and high-$Z$, few-electron ions have been reported using either Electron Beam Ion Trap (EBIT) or  Electron-Cyclotron Ion Sources (ECRIS) to produce ions at rest in the laboratory. Such measurements, using an EBIT, have been performed by the Livermore group (see, \eg \cite{mbvk1992,wbdb1996,bbbc2001,tbcc2008,bab2015} and reference there in), Heidelberg group \cite{bbkc2007,kbbl2012,rbes2013,esbr2015} and the Melbourne and National Institute of Standards and Technology (NIST) collaboration ~\cite{ckgh2012,cpgh2014,pckg2014}. The present collaboration has reported values using an ECRIS ~\cite{asgl2012}. 

The Heidelberg group reported the measurement of the $1s2p \,^1P_1 \rightarrow 1s^2 \,^1S_0$ He-like argon line with a relative accuracy of \num{1.5E-6} without the use of a reference line \cite{kbbl2012}. In that work, the spectrometer used is made of a single flat Bragg crystal coupled to a charge-coupled device (CCD) camera, which can be positioned very accurately with a laser beam reflected by the same crystal as the x~rays \cite{kbbl2012}. The Melbourne-NIST collaboration reported the measurement of all the $n=2\to n=1$ transitions in He-like titanium with a relative accuracy of \num{15E-6}, using a calibration based on neutral x-ray lines emitted from an electron fluorescence x-ray source \cite{ckgh2012,cpgh2014,pckg2014}. The Livermore group reported a measurement of all  $n=2\to n=1$ lines  in heliumlike copper  \cite{bab2015}, using hydrogenlike lines in argon as calibration. It also reported measurement of all 4 lines in He-like xenon, using a micro-calorimeter and calibration with x-ray standards \cite{tgbb2009}. It should be emphasized that  measurements in both type of ion sources do not require Doppler shift  correction  to transition energy measurements, because the ions have only thermal motion.

Measurements of the $1s 2s^2 2p \,^1P_1 \rightarrow 1s^2 2s^2\,^1S_0$ line in Be-like ions are scarce. Some measurements are relative measurements using tokamaks, where the Be-like line appears as a satellite line for the He-like $2\to1$ transitions. The  $1s2p \,^1P_1 \rightarrow 1s^2 \,^1S_0$  line is often  used as a calibration. Measurements of that type for Be-like  Ar have been performed at the \textit{Tokamak de Fontenay aux Roses} (TFR) \cite{tbbf1985}, and for Ni at the Tokamak Fusion Test Reactor (TFTR) \cite{bhzv1985,hbhv1987}. Such relative measurements, which use theoretical results on the He-like line, must be re-calibrated using the most recent theoretical values. Several other observations have been made on different elements, but no experimental energy reported (see, \eg Ref. \cite{rrag2015} for Cl, Ar and Ca), or the experimental accuracy is not completely documented (see \eg \cite{tcdl1985,rgtm1995,rrag2014}). Measurements in EBIT are also known, as in vanadium \cite{bcms1991} and iron \cite{bpjh1993}, for terrestrial  and astrophysics plasma applications. There have also been relative measurements in ECRIS for sulfur, chlorine and argon \cite{sbcs2013}, using the relativistic M1 transition  $1s2s \,^3S_1 \rightarrow 1s^2 \,^1S_0$ as a reference.

Chantler \etal~\cite{cak2009,ckgh2012,cpgh2014}, have claimed that existing data show the evidence of a  discrepancy between the most advanced BSQED calculation \cite{asyp2005} and  measurements in the He-like isoelectronic sequence, leading to a deviation that scales  as $\approx Z^3$. They speculated ~\cite{cpgh2014} that this supposed systematic effect could provide insight into the \textit{proton size puzzle}, the Rydberg and fine-structure constants, or missing three-body BSQED terms. Here we make a detailed analysis, including all available experimental results, to check this claim.

We emphasize the advantage of studying highly-charged, medium-$Z$ systems, such as argon ions, to test QED.  
The BSQED contributions have a strong $Z$-dependence: the retardation correction to the electron-electron interaction contribution scales as $Z^3$, and the one-electron corrections, self-energy and vacuum polarization, scale as $Z^4$. 
Yet, at high-$Z$, the strong enhancement of the nuclear size contribution and associated uncertainty limits the degree to which available experimental measurements can be used to test QED~\cite{pmgs1991,pas1995,pas1996,bmps1998,cak2009}. At very low-$Z$, experiments can be much more accurate, but tests of QED  can be limited as well, even for very accurate measurements of transitions to the ground state of He \cite{euvh1996,euvh1997,kgpu2011}. For few-electron atoms and ions, they are limited by the large size of electron-electron correlation and by the evaluation of  the needed higher-order QED screening corrections, in the non-relativistic QED formalism (NRQED) \cite{pac1998,pac1998a,pac2006,pac2006a,yap2010}. It can also be limited by the slow convergence of all-order QED contributions at low-$Z$, which may be required for comparison, and because of the insufficient knowledge of some nuclear parameters, namely the  form factors and  polarizability ~\cite{pana2010,ansa2013,pnfa2016,asyp2005}. In medium-$Z$ elements like argon or iron, the nuclear mean spherical radii are sufficiently well known (see, \eg \cite{aam2013}) and nuclear polarization contribution to the ion level energies is very small. So uncertainties related to the nucleus are small compared to experimental and theoretical  accuracy. This can be seen in the theoretical uncertainties claimed in Ref. \cite{asyp2005}.

Besides the fundamental aspect, knowledge of transition energies and wavelengths of highly-charged ions is very important for many sectors of research, such as astrophysics or plasma physics. For example, an unidentified line was recently detected in the energy range  \SI{3.55}{\keV} to \SI{3.57+-0.03}{\keV} in an X-ray Multi-Mirror (XMM-Newton) space x-ray telescope spectrum of 73 galaxy clusters\cite{bmfs2014} and at  \SI{3.52+-0.03}{\keV}  for another XMM spectrum in the Andromeda galaxy and the Perseus galaxy cluster \cite{brif2014}.  The next year a line at  \SI{3.539+-0.011}{\keV} was observed  in the deep exposure dataset of the Galactic center region with the same instrument. A possible connection with a dark matter decay line has been put forward, yet measurements performed with an EBIT seem to show that it could be a set of lines in highly charged sulfur ions, induced  by charge exchange \cite{sdbs2016}, while a recently published search with the high-resolution x-ray spectrometer of the HITOMI satellite does not find evidence for such lines in the Perseus cluster \cite{aaaa2017}.

In the present work, we apply the method we have developed to measure the energy and line-width of the $1s2s\,^3S_1\to1s^2 \,^1S_0$ M1 transition reported in Ref.~\cite{asgl2012}, to the $1s2p\,^1P_1\to1s^2 \,^1S_0$ transition in He-like argon and to the  $1s^2 2s 2p\,^1P_1\to1s^2 2s^2\,^1S_0$ transition in Be-like argon ions. We also present a multi-configuration Dirac-Fock (MCDF) calculation for the two transition energies and widths. These calculations are performed with a new version of the \textit{mcdfgme} code  that uses the effective operators developed by the St Petersburg group to evaluate the self-energy screening \cite{sty2013}.

The article is organized as follows. In the next section we briefly describe the experimental setup used in this work. A detailed description of the analysis method that provides the energy, width and uncertainties is given in Sec. \ref{sec:data-analysis}. 
A brief description of the calculations of transition energy and widths is given in Sec.~\ref{sec:mcdf}. 
We present our experimental result for the $1s 2p\,^1P_1\to 1s^2 \,^1S_0$ transition in Sec. \ref{sec:results-he}. In the same section we present all available experimental results for $7\leq Z\leq 92$ and $n=2 \to n=1$ transitions in He-like ions. We  do  a very detailed comparison between theory  from Ref. \cite{asyp2005}, which covers   $12\leq Z \leq 92$ and the available measurements in this $Z$-range. Our results and comparison with theory for the $1s 2s^2 2p\,^1P_1\to 1s^2 2s^2\,^1S_0$ line in  Be-like argon ions are presented in Sec. \ref{sec:results-be}. 
The conclusions are provided  in Sec.~\ref{sec:conclusions}.

%
\section{Experimental Method}
\label{sec:experimental}

ECRIS plasmas have been shown to be very intense sources of x~rays, and have diameters of a few cm. Therefore, they are better adapted to spectrometers that can use an extended source. At low energies one can thus use cylindrically or spherically bent crystal spectrometers as well as double-crystal spectrometers (DCSs). 

A single flat-crystal spectrometer, combined with an accurate positioning of the detector, and alternate measurements, symmetrical with respect to the optical axis of the instrument, as used in Heidelberg \cite{kbbl2012}, and the double-crystal spectrometers \cite{des1967,assg2014} are the only two methods that can provide high-accuracy, reference-free measurements in the x-ray domain. We use here \textit{reference-free} with the same meaning as in Ref. \cite{kdh1979}, \ie the measured wavelengths are directly connected to the meter as defined in the International System of Units, through the lattice spacing of the crystals \cite{assg2014}. Our group reported in 2012 such a measurement of the $1s2s \,^3S_1 \rightarrow 1s^2\, ^1S_0$ transition energy in He-like argon with an uncertainty of \num{2.5E-6} without the use of an external reference~\cite{asgl2012},  using the same experimental device  as in the present work: a DCS connected to an ECRIS, the ``Source d' Ions Multicharg\'es de Paris'' (SIMPA)\cite{gtas2010}, jointly operated by the Laboratoire Kastler Brossel and the Institute des Nanosciences de Paris on the Universit\'e Pierre and Marie Curie campus. 

A detailed description of the experimental setup of the DCS at the SIMPA ECRIS used in this work is given in Ref.~\cite{assg2014}. A neutral gas (Ar in the present study) is injected into the plasma chamber inside a magnetic system with minimum fields at the very center of the vacuum chamber. Microwaves at a frequency of \SI{14.5}{GHz} heat the electrons that are trapped by the magnetic field. The energetic electrons ionize the gas through repeated collisions reaching up to heliumlike charge states \cite{gasg2013}. The ions are, in turn,  trapped by the space charge of the electrons, which have a density around \SI{1e11}{\centi\meter^{-3}}. This corresponds to a trapping potential of a fraction of \SI{1}{\volt}, leading to an ion-speed distribution of $\approx$\SI{1}{\eV} per charge, and thus to a small Doppler broadening of all the observed lines. In contrast, EBITs have a trapping potential of several hundred \si{\eV}, and the Doppler broadening is then much larger.

The $1s2s \,^3S_1$ state is mostly created by electron ionization of the $1s ^2 2s \,^2S_{1/2}$ ground state of Li-like argon, and  therefore the $1s 2s \,^3S_{1} \to 1s^2 \, ^1S_0$ line is the most intense line we observed in He-like argon. The $1s 2p \,^1P_{1} \to 1s^2 \, ^1S_0$  line observed here results from the excitation of the $1s^2 \,^1S_0$ He-like argon ground state, which is much less abundant, leading to a weaker line. The Be-like excited level, $1s 2s^2 2p \,^1P_1$, is mostly produced  by ionization of the ground state of boronlike argon, which is a well-populated charge-state  (see Fig. 21, Ref. \cite{assg2014}).  The $1s 2s^2 2p \,^1P_1 \rightarrow 1s^2 2s^2\,^1S_0$ line is thus the most intense we observed.

The spectra are recorded by a specially-designed, reflection vacuum double-crystal spectrometer described in detail in Ref. \cite{assg2014}. The two $(6\times 4)$\si[product-units = power]{\centi\meter^2}, \SI{6}{\milli\meter}-thick Si(111) crystals were made at the National Institute for Standards and Technology (NIST). Their lattice spacing in vacuum was measured and found to be $d_{111}=$\SI{3.135 601 048(38)}{\angstrom} (relative uncertainty of \num{0.012E-6}) at a temperature of \SI{22.5}{\celsius} ~\cite{assg2014}, relative to the standard value \cite{fwz2011,mtn2012}. More details will be found in Ref. \cite{ksch2017}. Using this lattice spacing, our measurement provides wavelengths directly tied to the definition of the meter~\cite{mtn2012}. The DCS is connected to the ion source in such a way that the axis of the spectrometer is aligned with the ECRIS axis and is located at \SI{1.2}{\meter} from the  plasma (a sphere of $\approx$ \SI{3}{\cm} in diameter).

To analyze the experimental spectra, we developed a simulation code \cite{assg2014}, which uses  the geometry of the instrument and of the x-ray source, the shape of the crystal reflectivity profile, as well as the natural Lorentzian shape of the atomic line and  its Gaussian Doppler broadening to perform high-precision ray-tracing. 
The reflectivity profile is calculated using  XOP  (X-ray Oriented Programs) ~\cite{sad1998}, which uses dynamical diffraction theory from Ref.~\cite{zac1967}, and the result is checked with the X0H program, which calculates crystal susceptibilities $\chi_0$ and $\chi_h$~\cite{stepanov,las1991}.

The first crystal is maintained at a fixed angle. A spectrum is obtained by a series of scans of the second crystal. A stepping motor, driven by a micro-stepper, runs continuously, between two predetermined angles that define the angular range of one spectrum. X~rays are recorded continuously and stored in a histogram, together with both crystals temperatures. Successive spectra are recorded in opposite directions. Both crystal angles are measured with Heidenhain\footnote{Certain commercial equipment, instruments, or materials are identified in this paper to foster understanding. Such identification does not imply recommendation or endorsement by the National Institute of Standards and Technology, nor does it imply that the materials or equipment identified are necessarily the best available for the purpose.} high-precision angular encoders.
The experiment is performed in the following way: a nondispersive-mode (NDM) spectrum is recorded first. Then a dispersive-mode (DM) spectrum is recorded. The sequence is completed with the recording of a second NDM  spectrum. Due to the low counting rate,  such a sequence of three spectra  takes a full day to record. In order to obtain enough statistics, the one-day sequence is repeated typically  7 to 15 times.

\section{Data analysis}
\label{sec:data-analysis}
The data analysis is performed in three steps. First we derive a value for the experimental natural width of the line. For this, each experimental dispersive-mode spectrum is fitted with simulated spectra, using an approximate energy (\eg the theoretical value) and a set of Lorentzian widths. A weighted one-parameter fit is performed on all the results for  all recorded dispersive-mode spectra providing a width value and its uncertainty. This experimental width is then used to generate a new set of simulations, using several different energies and crystal temperatures. These simulations are used to fit each dispersive-mode  and nondispersive-mode experimental spectrum in order to obtain the line energy.  For each day of data recording this leads to two Bragg angle values, obtained by taking the angular difference between a nondispersive-mode spectrum and a dispersive-mode spectrum:
\begin{itemize}
\item one Bragg angle value is obtained by comparing the first nondispersive-mode spectrum of the day and the dispersive-mode spectrum obtained immediately after;
\item a second Bragg angle value is obtained by comparing  the same dispersive-mode spectrum with the nondispersive-mode spectrum obtained immediately after.
\end{itemize}
In that way a number of possible time-dependent drifts in the experiment are compensated. We now describe these processes in more detail.

\subsection{Evaluation of the widths}
\label{subsec:widths}

The ion temperature, which is necessary to calculate the Gaussian broadening was obtained by measuring first a line with a completely negligible natural width, the  M1 $1s2s\, ^3S_1 \rightarrow 1s^2\, ^1S_0$, transition. The  width of this transition is \SI{\approx 1e-7}{\eV}, which is totally negligible when compared to our spectrometer inherent energy resolution.  From this analysis we obtained the Gaussian broadening $\Gamma_{\textrm{G}}^{\textrm{Exp.}}=$\SI{80.5\pm4.6}{\meV} \cite{asgl2012}. This value also provides the depth of the trapping potential due to the electron space charge. Knowing the experimental Gaussian broadening value {$\Gamma_{\textrm{G}}^{\textrm{Exp.}}$,  we can perform all the needed simulations. 
For  each line under study we then proceed as follows:
\begin{itemize}
\item Perform simulations for the dispersive-mode spectra for a set of natural width values $\Gamma^i_{\textrm{L}}$ and  the theoretical transition energy $E_0$ , using the already known 
$\Gamma_{\textrm{G}}^{\textrm{Exp.}}$, and crystal temperature $T_{\textrm{Ref.}}=\,$\SI{22.5}{\degreeCelsius};

\item Interpolate each simulation result with a piece-wise spline function to obtain a set of continuous, parametrized functions $S_{\left[E_0,\Gamma^i_{\textrm{L}},\Gamma_{\textrm{G}}^{\textrm{Exp.}},T\right]}\left(\theta-\theta_0\right)$, where $\theta_0$ correspond to the angle at which the simulation reaches its maximum value, and $T=T_{\textrm{Ref.}}$;

\item Normalize all the functions above to have the same maximum value (we chose the one with $\Gamma_{\textrm{L}}=0$ as reference);

\item Fit each experimental spectrum with  the functions obtained above
\begin{equation}
I\left(\theta-\theta_0,I_{\textrm{max}},a,b\right)=I_{\textrm{max}} S_{\left[E_0,\Gamma^i_{\textrm{L}},\Gamma_{\textrm{G}}^{\textrm{Exp.}},T\right]}\left(\theta-\theta_0\right)+ a + b \theta,
\label{eq:fit_func}
\end{equation}
where $I_{\textrm{max}}$ is the line intensity,  $\theta$ the crystal angle, $a$ the background intensity and $b$ the background slope. The parameters $\theta_0$, $I_{\textrm{max}}$, $a$ and $b$ are adjusted to minimize the  reduced $\chi^2\left(\Gamma^i_{\textrm{L}}\right)$. We perform a series of fits of each experimental spectrum, with 27 simulated spectra, each evaluated with a different width  $\Gamma^i_{\textrm{L}}$, to obtain a set of  $\chi^2\left(\Gamma^i_{\textrm{L}}\right)$ values. The width values go from  \SI{0}{\milli\eV} to \SI{250}{\milli\eV} by steps of \SI{10}{\milli\eV}, completed by a point at \SI{300}{\milli\eV}. A typical experimental spectrum and the fitted simulated functions, for 5 of the 27  values of $\Gamma^i_{\textrm{L}}$ used to make the analysis, are shown in Fig. \ref{fig:he-width-ex} ;

\item  Fit a third degree polynomial to the set of points $[\Gamma^i_{\textrm{L}},\chi^2\left(\Gamma^i_{\textrm{L}}\right)]$;

\item Find the minimum of the third degree polynomial to get the corresponding optimal $\Gamma_{\textrm{L\,opt.}}^{n}$, $n$ being the experiment run number (see Fig. \ref{fig:with-chi2} for an example);

\item Get the \SI{68}{\percent} error bar $\delta \Gamma_{\textrm{L\,opt.}}^{n}$  for experiment run $n$ by finding the values of the width for which \cite{pftv2007}
\begin{equation}
\chi^2\left(\Gamma_{\textrm{L\,opt.}}^{n}\pm \delta \Gamma_{\textrm{L\,opt.}}^{n}\right)=\chi^2\left(\Gamma_{\textrm{L\,opt.}}^{n}\right)+1;
\label{eq:error-width}
\end{equation}
\item Finally a weighted average of the values in the set of all the $\Gamma_{\textrm{L\,opt.}}^n$ obtained for all measured spectra is performed to obtain the experimental value $\Gamma_{\textrm{L}}^{\textrm{Exp.}}$ and its error bar:
\begin{eqnarray}
\label{eq:gamma_ave}
\frac{1}{\left(\delta \Gamma_{\textrm{L}}^{\textrm{Exp.}}\right)^2}&=& \sum_n \frac{1}{\left(\delta\Gamma_{\textrm{L\,opt.}}^{n}\right)^2} \, ,\nonumber \\
\Gamma_{\textrm{L}}^{\textrm{Exp.}}&=& \left(\delta \Gamma_{\textrm{L}}^{\textrm{Exp.}}\right)^2\sum_n \frac{\Gamma_{\textrm{L\,opt.}}^n}{\left(\delta\Gamma_{\textrm{L\,opt.}}^{n}\right)^2} \, .
\end{eqnarray}
The sets of $\Gamma_{\textrm{L\,opt.}}^n$ for both lines studied here are plotted in Fig. \ref{fig:wv}. 

\end{itemize}
The two first steps are performed by two different methods, one based on the CERN (Centre Européen de Recherche Nucléaire) program ROOT, version 6.08 ~\cite{bar1997,abbb2007,abbb2011} and one based on MATHEMATICA, version 11 ~\cite{mathematica11}. 

%
\

%
\begin{figure}
\centering
\includegraphics[clip=true,width=13cm]{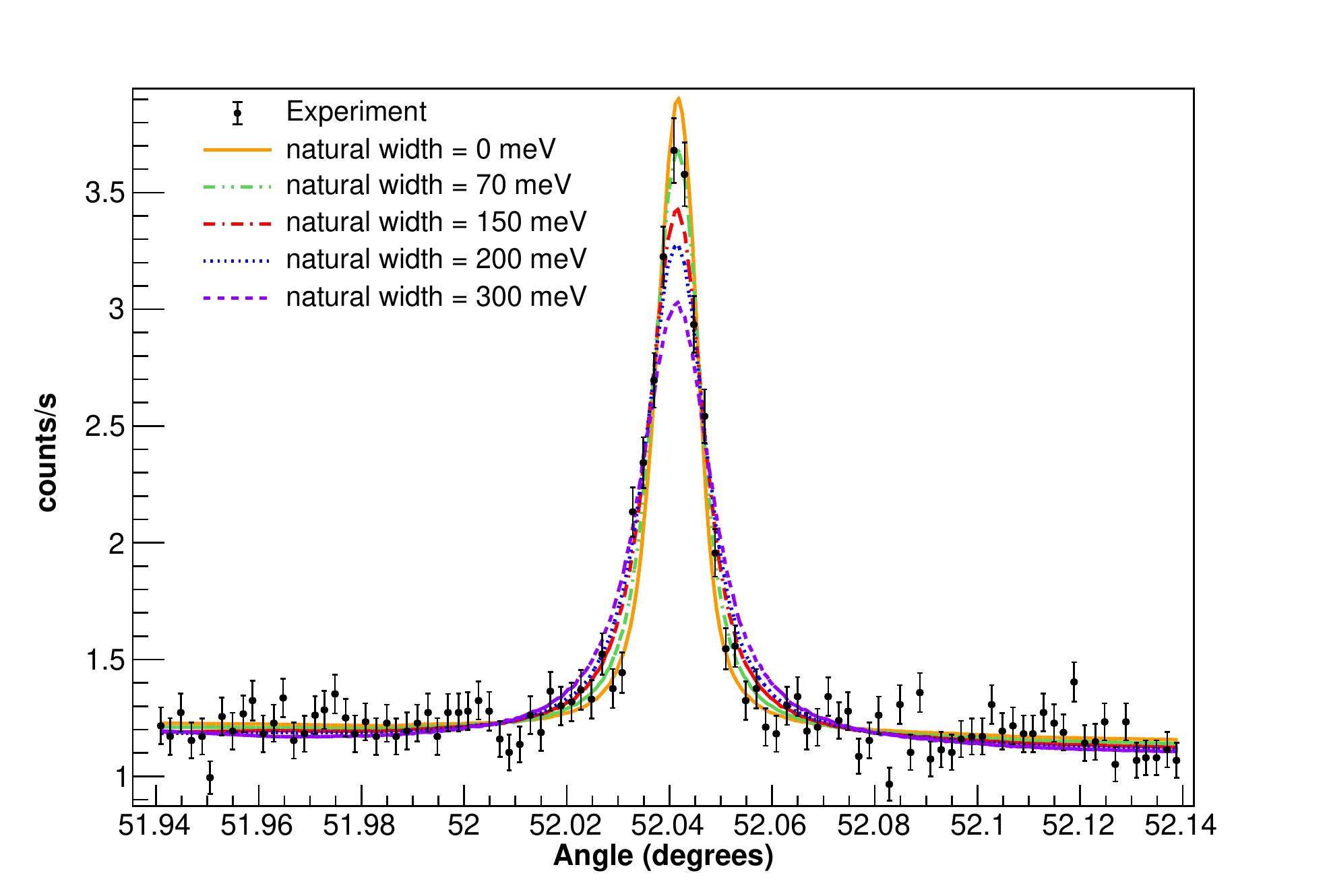}
\caption{
(Color online) Example of a dispersive-mode experimental spectrum for the He-like Ar $1s2p \,^1P_1 \rightarrow 1s^2\,^1S_0$ transition (black dots), together with a few plots of the function in Eq. \eqref{eq:fit_func},
for different values of the natural line width  $\Gamma_{\textrm{L}}^i$. The four parameters have been adjusted to minimize the reduced $\chi^2\left(\Gamma_{\textrm{L}}\right)$ (see text for more explanations).
\label{fig:he-width-ex}
}
\end{figure}
%
\

%
\begin{figure}
\centering
\includegraphics[clip=true,width=13cm]{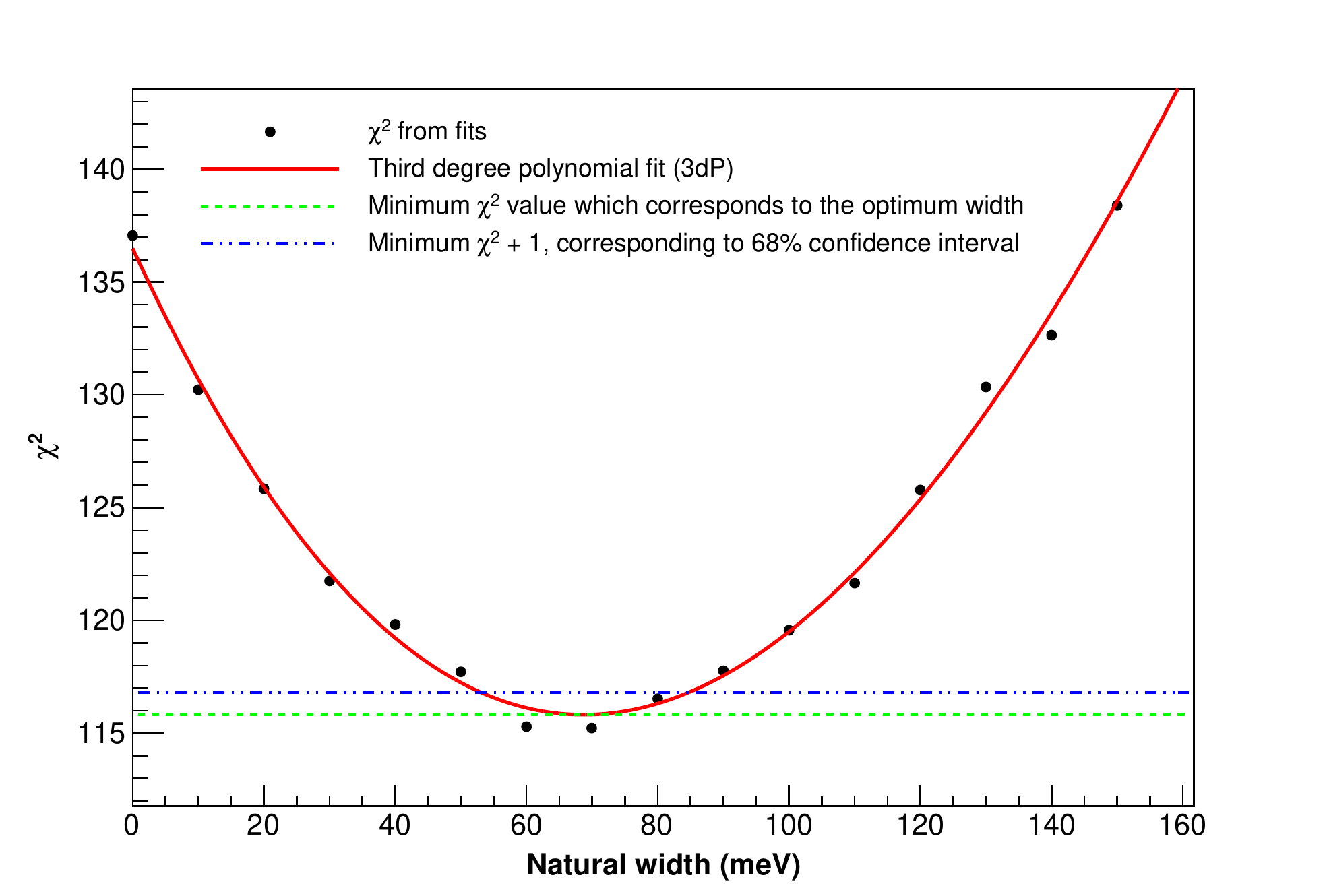}
\caption{
 (Color online) Third degree polynomial fitted to the $[\Gamma_{\textrm{L}},\chi^2\left(\Gamma_{\textrm{L}}\right)]$ set of points (black dots), for the He-like Ar 1s2p $^1P_1 \rightarrow$1s$^2$ $^1S_0$ transition. The $\chi^2$ values were obtained from the fits, a few of which are represented in Fig. \ref{fig:he-width-ex}, with 27 different values of $\Gamma_{\textrm{L}}$. The blue dashed-doted line corresponds to Eq. \ref{eq:error-width}.
\label{fig:with-chi2}
}
\end{figure}
%
\begin{figure}
    \centering
    \begin{subfigure}[a]{0.65\textwidth}
        \includegraphics[width=\textwidth]{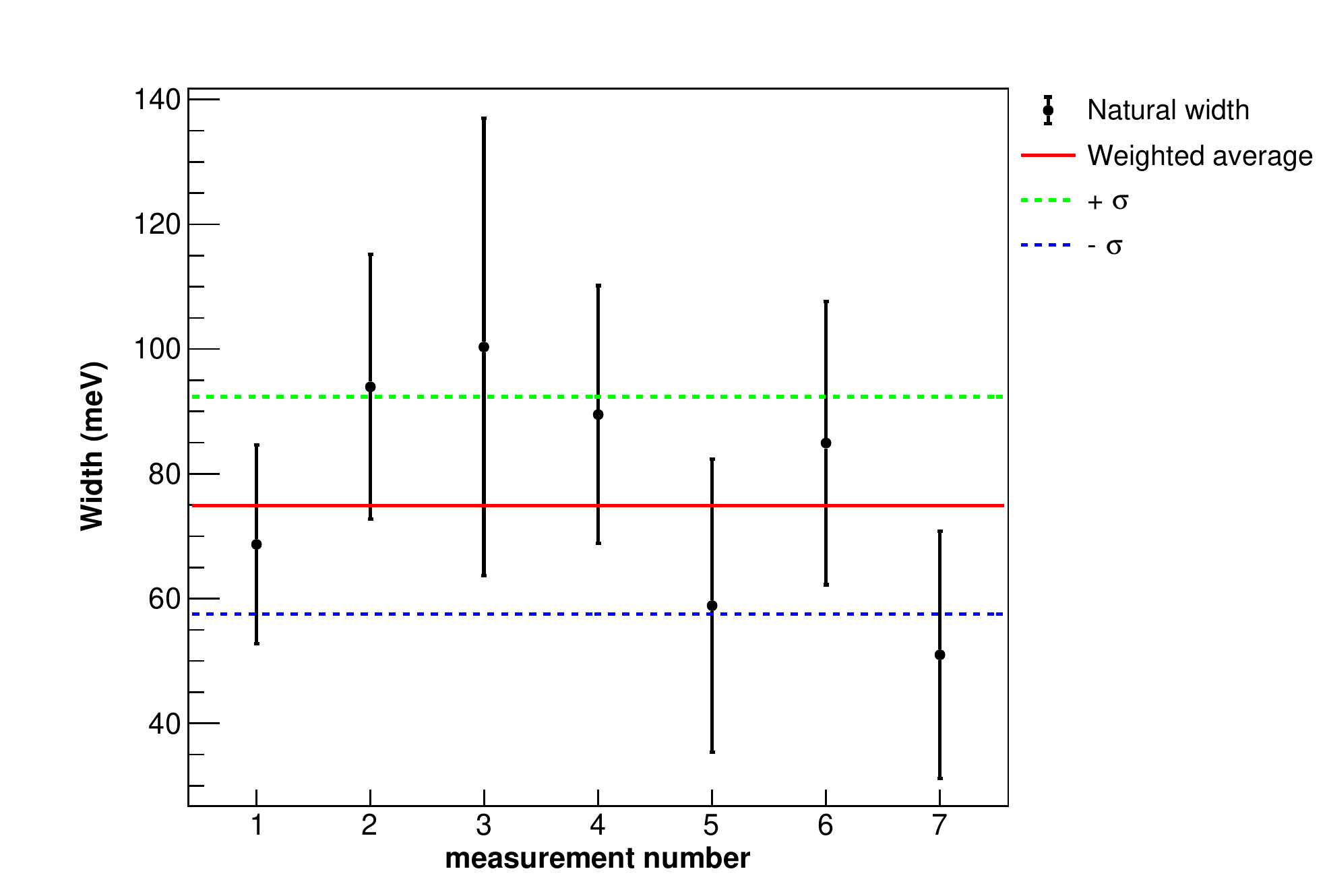}
        \caption{He-like argon $1s2p \,^1P_1 \rightarrow 1s^2\,^1S_0$ transition.}
        \label{fig:whe-like}
    \end{subfigure}
    
    ~ 
    \begin{subfigure}[b]{0.65 \textwidth}
        \includegraphics[width=\textwidth]{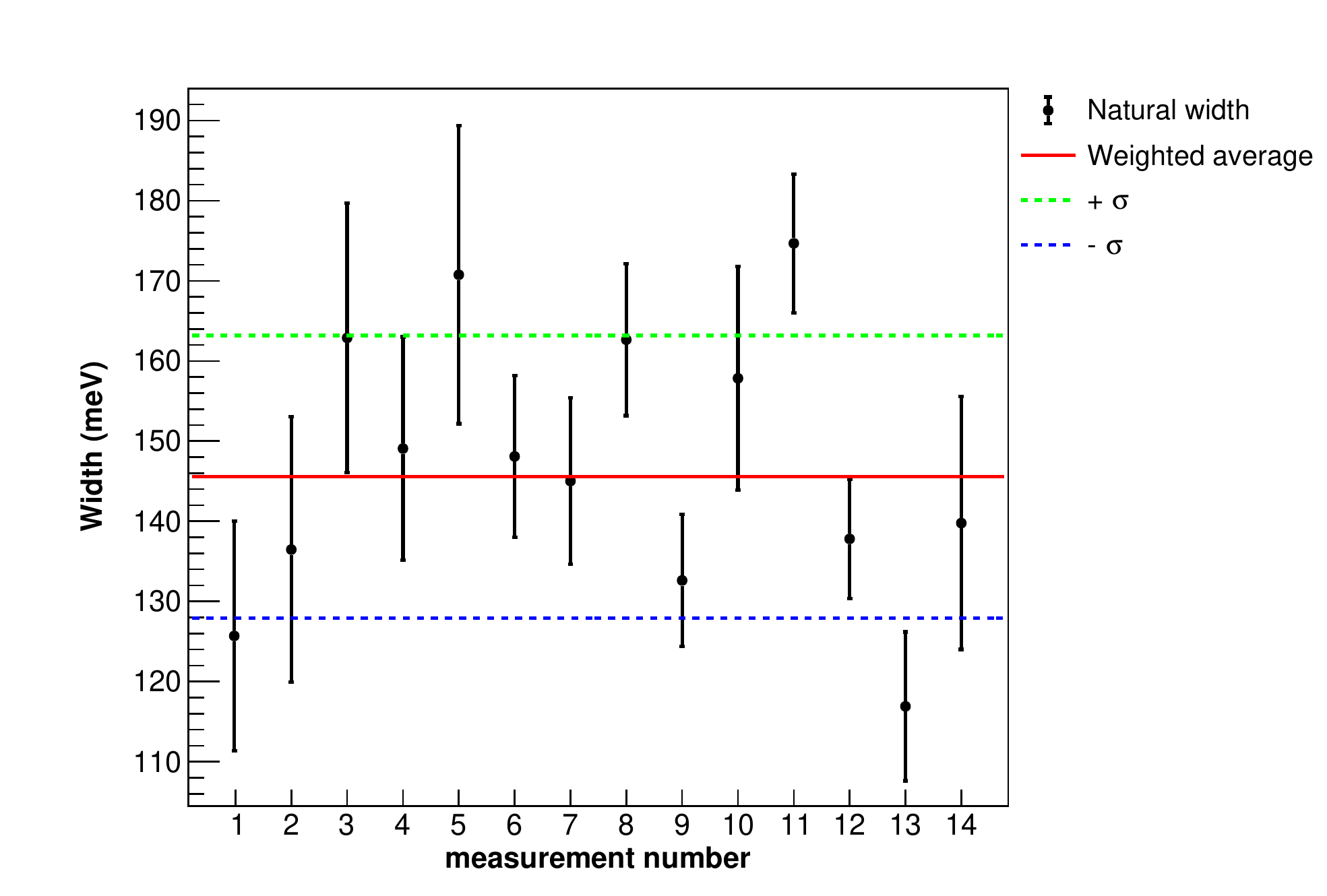}
        \caption{Be-like argon $1s 2s^2 2p \,^1P_1 \rightarrow 1s^2 2s^2\,^1S_0$ transition.}
        \label{fig:wbe-like}
    \end{subfigure}
    \caption{(Color online) Natural width values of all the spectra recorded during the experiment, with weighted average and uncertainties  evaluated with Eq. \eqref{eq:gamma_ave}.}
    \label{fig:wv}
\end{figure}

\subsection{Transition energy values}
\label{sec:tev}

Once we obtained the experimental width value  $\Gamma_{\textrm{L}}^{\textrm{Exp.}}$ of a measured line (cf. Sec. \ref{subsec:widths}), the determination of the correspondent experimental transition energy value $E_\textrm{exp}$ is achieved using the following scheme:

\begin{itemize}
\item Perform simulations in the nondispersive  and dispersive modes for a set of transition energy values $E_k=E_\textrm{theo} + k\Delta E$, where $E_\textrm{theo}$ is the theoretical energy value,  $ \Delta E$ an energy increment and $k$ an integer that can take positive or negative values. The simulations are done with the experimental natural width  $\Gamma_{\textrm{L}}^{\textrm{Exp.}}$ and Gaussian broadening $\Gamma_{\textrm{G}}^{\textrm{Exp.}}$. The simulations are performed at various crystal temperature values $T_{l}$ for each energy. 

\item As in Sec. \ref{subsec:widths}, interpolate each simulation result with a spline function for both the  nondispersive and dispersive modes, to obtain a set of functions depending on all the $\left(E_k, T_l\right)$ pairs;

\item Fit each experimental spectrum, using Eq. \eqref{eq:fit_func} with $E_0=E_k$ and $T=T_l$, to obtain the angle difference between the simulation and the experimental spectrum, both in dispersive and nondispersive mode;

\item For each pair of dispersive  and nondispersive modes experimental spectra, calculate the offsets   $\Delta \theta_{\textrm{Exp.}-\textrm{Simul.}}^{n,k,l}= \left(\theta_{\textrm{Exp.}\textrm{DM}}^n -\theta_{\textrm{Exp.}\textrm{NDM}}^n\right)- \left(\theta_{\textrm{Simul.}\textrm{DM}}^{k,l} -\theta_{\textrm{Simul.}\textrm{NDM}}^{k,l}\right)$ between the simulated spectra and the experimental value  obtained in the step above. This offset should be 0 if the energy and temperature used in the simulation were identical to the experimental values;

\item Fit the bidimensional function 
\begin{equation}
\Delta \theta_{\textrm{Exp.}-\textrm{Simul.}}(E,T)=p+qE +rE^2+ sET + uT + vT^2,
\label{eq:enertemp}
\end{equation}
where $p,q,r,s,u$ and $v$ are adjustable parameters, to the set of points $\left[E_k,T_l,\Delta \theta_{\textrm{Exp.}-\textrm{Simul.}}^{n,k,l}\right]$ obtained in the previous step (see Fig. \ref{fig_2dfhl} as an example);

\item The experimental line energy $E_\textrm{Exp.}^n$ for spectrum pair number $n$, is the energy such that $\Delta \theta_{\textrm{Exp.}-\textrm{Simul.}}\left(E_\textrm{Exp.}^n,T_\textrm{Exp.} \right)=0$ where $T_\textrm{Exp.}$, stands for the average measured temperature on the second crystal;

\item As a check, we also used the line energy such that $\Delta \theta_{\textrm{Exp.}-\textrm{Simul.}}\left(E_\textrm{Exp.}^n,T_\textrm{Ref.} \right)=0$ ($T_\textrm{Ref.} =$\SI{22.5}{\celsius}). This leads to a temperature-dependent energy. We then fitted a straight line to the line energy, as a function of the second crystal temperature, and extrapolated to $T=$\SI{22.5}{\celsius}. Both methods lead to very close values, well within the uncertainties.

\item As in Sec. \ref{subsec:widths}, we calculate  the weighted average of all the  $\left(n,E_\textrm{Exp.}^n\right)$ pairs to obtain the final experimental energy. The error bar on each point is the quadratic combination of the instrumental uncertainty, as given in Table \ref{tab:errors} and of the statistical error. 

\item To check the result, we also fit the set of $\left(E_\textrm{Exp.}^n,T_\textrm{Exp.}^n \right)$ pairs  with the function $E_0 + b T$ to check that there is no residual temperature dependence. 

\end{itemize}

\begin{table}[tb]
\begin{center}
\begin{ruledtabular}
\caption{
Instrumental contributions to the uncertainties in the analysis of the daily experiments (see Refs. \cite{asgl2012,assg2014}).}
\label{tab:errors}
\begin{tabular}{lD{.}{.}{10}}
 \multicolumn{1}{c}{Contribution: }& \multicolumn{1}{c}{Value (eV) }\\
\hline
Crystal tilts (\SI{\pm 0.01}{\degree} for each crystal)	&	0.0002	\\
Vertical misalignment of collimators (\SI{1}{\mm})	&	0.0002	\\
X-ray source size (\SI{6}{\mm} to \SI{12}{\mm})	&	0.0013	\\
\hline							
Form factors 	&	0.0020	\\
X-ray polarization	&	0.0014	\\
\hline							
Angle encoder error	&	0.0036	\\
Lattice spacing error	&	0.00012	\\
Index of refraction	&	0.0016	\\
Coefficient of thermal expansion	&	0.00019	\\
X-ray polarization	&	0.00100	\\
Energy-wavelength correction	&	0.000078	\\
Temperature (\SI{0.5}{\celsius})	&	0.0040	\\
\end{tabular}
\end{ruledtabular}
\end{center}
\end{table}
%

%
\begin{figure}
\centering
\includegraphics[clip=true,width=13cm]{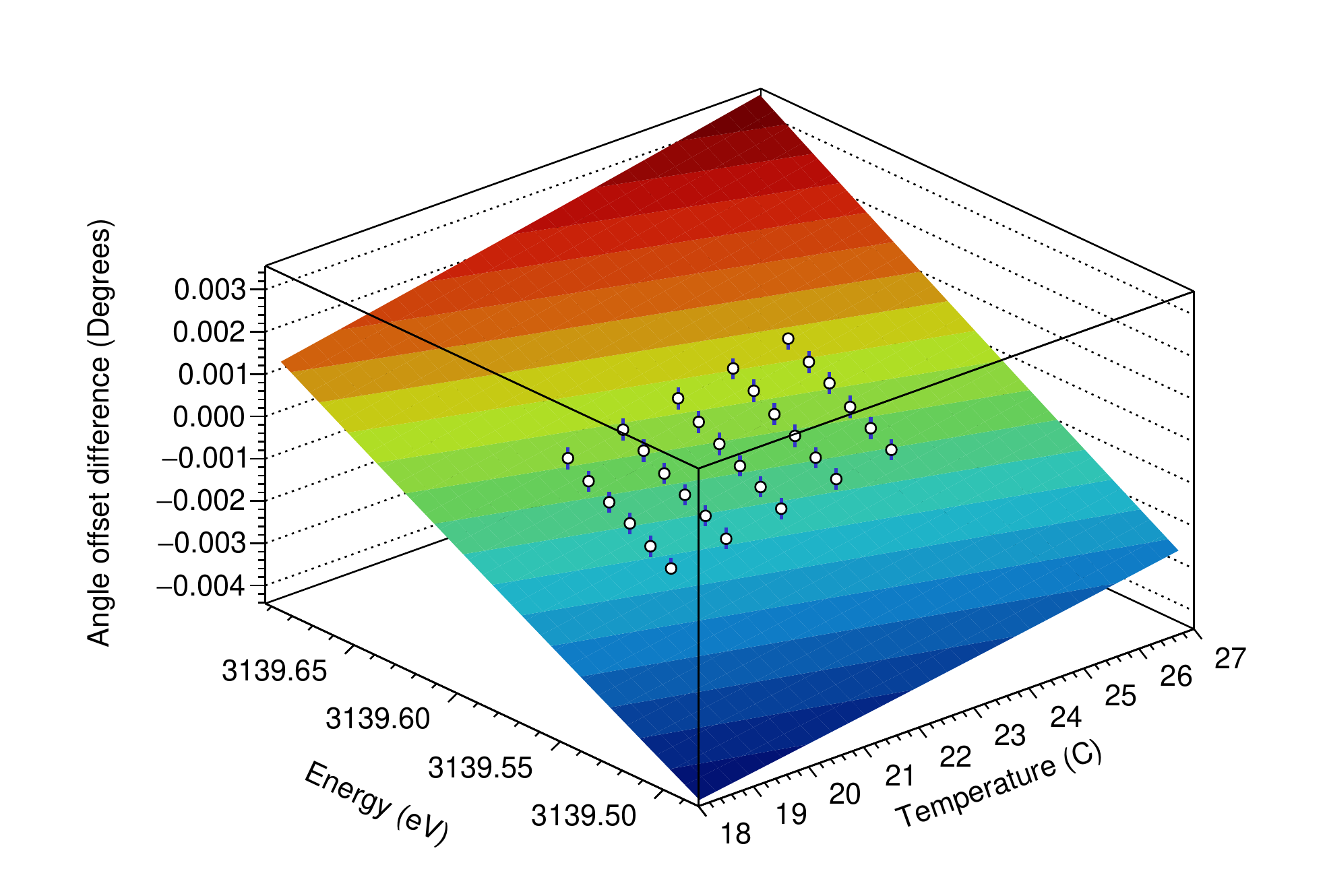}
\caption{
(Color online) Fitted two-dimensional function from Eq. \eqref{eq:enertemp}, and  experimental results (white  spheres), for the He-like Ar $1s2p \,^1P_1 \rightarrow 1s^2\,^1S_0$ transition. The fit is performed taking into account the  statistical error bars in each point.
\label{fig_2dfhl}
}
\end{figure}
%

%
%
%
%
%
%

\section{Theoretical calculation}
\label{sec:mcdf}

The core-excited $1s 2s^2 2p \,^1P_1 \rightarrow 1s^2 2s^2\,^1S_0$ transition in Be-like ions has been calculated with the most recent methods, only very recently and only for iron \cite{ysf2014}, and argon \cite{ysf2015}. Previous calculations \cite{sau1979,sal1979,che1985,cac1987} did not take into account QED and relativistic effects to the extent possible today.

For the preparation of this experiment, we performed a calculation of the  energy value for the $1s 2s^2 2p \,^1P_1 \rightarrow 1s^2 2s^2\,^1S_0$ transition in Be-like argon, using the multiconfiguration Dirac-Fock (MCDF) approach as implemented in  the 2017.2 version of the  relativistic MCDF code (MCDFGME),  developed by Desclaux and Indelicato~\cite{des1975,iad1990,igd1987,iad2005}. The full description of the method and the code can be obtained from Refs.~\cite{gra1970,des1975,gac1988,ind1995}. The present version also takes into account the normal and specific mass shifts, evaluated following the method of Shabaev \cite{sha1985,saa1994,sha1998}, as described in \cite{lngf2012,spng2013}. 

The main advantage of the MCDF approach is the ability to include a large amount of electronic correlation by taking into account a limited number of configurations~\cite{smpl1999,srmp2006,mmcs2009}.
%
All calculations were done for a finite nucleus using a uniformly charged sphere. The atomic masses and the nuclear radii were taken from the tables by Audi \etal~\cite{awt2003} and Angeli and Marinova~\cite{ang2004,aam2013}, respectively. 

Radiative corrections are introduced from a full QED treatment. The one-electron self-energy is evaluated using the one-electron values of Mohr and co-workers ~\cite{moh1974,moh1992,mak1992, iam1992,lim2001}, and corrected for finite nuclear size~\cite{mas1993}. The self-energy screening and vacuum polarization were included using the methods developed by Indelicato and co-workers~\cite{igd1987, iad1990, ial1992, ibl1998, iam1998}. In previous work, the self-energy screening in this code was based on the Welton approximation \cite{igd1987,iad1990}. Here we also evaluate the self-energy screening following the model operator approach recently developed by Shabaev \etal \cite{sty2013,sty2015}, which has been added to MCDFGME. A detailed description of this new code will be given elsewhere. 

In order to assess the quality of this new method for calculating the self-energy screening we can compare the different values for the He-like transition measured here. The QED value of Indelicato and Mohr \cite{iam2001} is \SI{0.1100}{\eV}, the one from Ref. \cite{asyp2005} (Table IV) is \SI{0.1085}. The Welton method provides \SI{0.0916}{\eV}, while the implementation of the Saint-Petersburg effective operator method gives \SI{0.0965}{\eV}, closer to the ab initio methods. We can thus assume an uncertainty of \SI{0.014}{\eV} and \SI{0.018}{\eV} for the effective operator and Welton operator methods respectively. The same procedure applied to the Be-like transitions provides \SI{0.130}{\eV} using Ref. \cite{iam2001}, \SI{0.112}{\eV} for the effective operator method and \SI{0.109}{\eV} for the Welton method. We can conclude that at intermediate $Z$, both the Welton and effective operator methods provide very similar results, the effective operator method being in slightly better agreement with ab initio calculation. This is consistent with earlier comparisons for fine-structure transitions, (see, \eg, Ref. \cite{blu1993}).

Lifetime evaluations are done using the method described in Ref. \cite{ipm1989}. The orbitals contributing to the wave function were fully relaxed, and the resulting non-orthogonality between initial and final wave functions  fully taken into account, following \cite{low1955,ind1997}. 

The full Breit interaction and the Uehling potential are included in the self-consistent field process. Projection operators have been included \cite{ind1995} to avoid coupling with the negative energy continuum. 

As a check, we also performed   a calculation of the He-like argon lines measured in the present work and in Ref. \cite{asgl2012}.  Following Refs. \cite{fro1977,gid1987,ind1995,ind1996}, we use for the excited state the following configurations:
\begin{eqnarray}
\left| 1s 2p\,^1P_1\right\rangle&=& c_1 \left|1s   2p, J=1 \right\rangle+c_2 \left|2s   3p, J=1 \right\rangle+c_3 \left|2p' 3d, J=1 \right\rangle+c_4 \left|3s   4p, J=1 \right\rangle \nonumber \\
                                               &  & +c_5 \left|3p'  4d, J=1 \right\rangle+c_6 \left|3d'  4f, J=1 \right\rangle+c_7 \left|4s   5p, J=1 \right\rangle+c_8 \left|4p'  5d, J=1 \right\rangle \nonumber \\
                                               &  & +c_9 \left|4d'  5f, J=1 \right\rangle+c_{10} \left|4f' 5g, J=1 \right\rangle+c_{11} \left|5s   6p, J=1 \right\rangle+c_{12} \left|5p' 6d, J=1 \right\rangle\nonumber \\
                                               &  & +c_{13} \left|5d'  6f, J=1 \right\rangle+c_{14} \left|5f' 6g, J=1 \right\rangle+c_{15} \left|5g' 6h, J=1 \right\rangle,
\end{eqnarray}
where the $l'$ indicates an orbital with identical angular function as the $l$ one, but with another radial wave function, for which the orthogonality with orbitals of the same symmetry in other configuration is not enforced. The  ground state wave function  is taken as usual as $\left| 1s^2\,^1S_0\right\rangle=c_1 \left| 1s^2, J=0\right\rangle+ c_2 \left| 2s^2, J=0\right\rangle+c_3 \left| 2p^2, J=0\right\rangle+\cdots+c_{20} \left| 6g^2, J=0\right\rangle+c_{21} \left| 6h^2, J=0\right\rangle$. We also evaluated  
\begin{eqnarray}
\left| 1s 2s\,^3S_1\right\rangle&=& c_1 \left|1s   2s, J=1 \right\rangle+c_2 \left| 2p   3p, J=1 \right\rangle+c_3 \left| 3s 4s J=1 \right\rangle+c_4 \left|3d   4d, J=1 \right\rangle \nonumber \\
                                               &  & +c_5 \left|4p  5p, J=1 \right\rangle+c_6 \left|4f  5f, J=1 \right\rangle+c_7 \left|5s   6s, J=1 \right\rangle+c_8 \left|5d  6d, J=1 \right\rangle \nonumber \\
                                               &  & +c_9 \left|5g  6g, J=1 \right\rangle,
\end{eqnarray}
in order to calculate the M1 transition energies measured Ref. \cite{asgl2012}, which allowed to compare also energy differences.

For Be-like argon, the correlation contributions result from the inclusion of all single, double and triple electron excitations of the $n=1$ and 2 electrons in the unperturbed configuration up to $n=5$.  
For the $1s^2 2s^2\,^1S_0$ ground state it corresponds to \num{2478} configurations and for the $1s 2s^2 2p \,^1P_1$ excited state to \num{14929} configurations.  We performed an estimation of the full correlation energy by doing a fit with the function $ a + b/n^2 + c/n^3$, and extrapolation to $n\to \infty$ for each level, for both the Welton and  the Model operator values. The results are presented in Table \ref{tab:be-like-total-energy}. By comparing the extrapolated value and the changes in QED due to the use of either the Welton or effective operator method we estimated 
the theoretical uncertainty provided in the table. There is however a contribution that is not included, the Auger shift. This shift is due to the fact that the  $1s 2s^2 2p \,^1P_1$ being core-excited is degenerate with a continuum. To our knowledge, such shifts have been evaluated only in the case of neutral atoms x-ray spectra \cite{ial1992,ibl1998,dkib2003}. For argon with a $1s$ hole, the shift is \SI{165}{\meV}, while for a $2p$ hole it is \SI{11}{\meV}. Here we have  a 4-electron system, with only 3 possible Auger channels, and the $2s$ shell is closed, so the effect is expected to be small. We assume an extra theoretical uncertainty of  \SI{11}{\meV} for this uncalculated term.

\begingroup
\squeezetable							
\begin{table*}							
\caption{							
Total  energy  and transition energies ( in \si{\eV})  for the  $1s 2s^2 2p \,^1P_1 \rightarrow 1s^2 2s^2\,^1S_0$ transition in Be-like argon, as a function of the maximum principal quantum number $n$ of the correlation orbitals.
All correlation from the Coulomb, retardation and QED parts is included.	Extrapolation for $n\to \infty$ is done by fitting the function $ a + b/n^2 + c/n^3$ to the correlation energy (Difference with the energy for $n$ and the Dirac-Fock (DF) value) of each level and retaining only the constant term $a$. The uncertainty combines the difference between the extrapolated and best directly calculated value, the missing Auger shift and the self-energy screening model.
}
\label{tab:be-like-total-energy} 							
\begin{ruledtabular}
\begin{tabular}{c|ddd|ddd}	
	&	\multicolumn{3}{c|}{Welton QED}					&	\multicolumn{3}{c}{Model operator QED\cite{sty2013,sty2015}}					\\
$n$	&	\multicolumn{1}{c}{Initial}	&	\multicolumn{1}{c}{Final}	&	\multicolumn{1}{c|}{Transition}	&	\multicolumn{1}{c}{Initial}	&	\multicolumn{1}{c}{Final}	&	\multicolumn{1}{c}{Transition}	\\
	\hline
\multicolumn{1}{c|}{DF} & -7222.7485 & -10313.5817 & 3090.8333 & -7222.7522 & -10319.3215 & 3096.5692\\

2 & -7227.3514 & -10319.3250 & 3091.9736 & -7227.3551 & -10319.3320 & 3091.9769\\

3 & -7228.6879 & -10320.5341 & 3091.8462 & -7228.6915 & -10320.5417 & 3091.8502\\

4 & -7229.0470 & -10320.7556 & 3091.7086 & -7229.0506 & -10320.7638 & 3091.7131\\

5 & -7229.1988 & -10320.8783 & 3091.6795 & -7229.2024 & -10320.8870 & 3091.6846\\
$\infty $ & -7229.4027 & -10321.1125 & 3091.7098 & -7229.4064 & -10321.1225 & 3091.7161 \\ 
\end{tabular}							
\end{ruledtabular}
\end{table*}																			
\endgroup

The Auger width of the $1s 2s^2 2p \,^1P_1 $ level is calculated with the MCDFGME code, following the method described in Ref. \cite{hag1978} with full relaxation and final-state channel mixing, again taking into account the non-orthogonality between the initial and final state.  For the first time, we combine this method with fully correlated wave functions, up to $n=5$. The convergence of the transition energy and width are presented in Table  \ref{tab:be-rate-conv}. This table shows that the Auger width values vary rather strongly when increasing the maximum $n$ of correlation orbitals, when non-orthogonality and full relaxation are included. This behavior is due to the fact that the free electron wave functions have to be orthogonal to all the occupied and correlation  orbitals of the same symmetry, which provides a lot of constraints.  

We have also performed calculations of the transition energies and rates with the ``flexible atomic code'' (FAC),  widely used in plasma physics \cite{gu2008}. This code is based on the relativistic configuration interaction (RCI), with independent particle basis wave functions that are derived from a local central potential. This local potential is derived self-consistently to include the screening of the nuclear potential by the electrons.

The final results are compared to other calculations from Refs. \cite{che1985,cmps2001,nat2003} in Table \ref{tab:be-rate-theory}. The relatively large difference between our present MCDF calculation and the Dirac-Fock calculation from Ref. \cite{cmps2001} , made with an earlier version of our code, is due to correlation and to the evaluation of Auger rates using fully relaxed initial and final states. 

The contributions of all the other possible transitions to the $1s^2 \, nl \,J$ levels, $n=3 \to \infty$, was evaluated by computing all Auger widths up to $n=9, l=8$.  
We then fitted  a function $a/n^2+b/n^3$ to the total  Auger width for each principal quantum number $n$, summing all values of $L$ and $J$ for each value of $n$, to evaluate the contribution from $n=10$ up to infinity. We find $a=$\SI{0.0562325}{\meV} and $b=$\SI{0.53028}{\meV}. The total value for the contribution of all levels with $n\ge 3$ is \SI{0.063}{\meV} and is thus negligible.

\begingroup
\squeezetable							
\begin{table*}							
\caption{							
Convergence of theoretical partial radiative widths,  Auger widths and  energies for transitions originating from the Be-like  $1s 2s^2 2p\, ^1P_1$ level. Transition energies are in \si{\eV} and  widths in  \si{\meV}. 	
}
\label{tab:be-rate-conv} 							
\begin{ruledtabular}
\begin{tabular}{c|dd|dd|dd|dd|d}	
	&	\multicolumn{2}{c|}{Radiative}  			&	\multicolumn{6}{c|}{Auger}  	&					\\
	\hline
	&	\multicolumn{2}{c|}{$\to1s^2 2s^2 \,^1S_0$}  			&	\multicolumn{2}{c|}{$\to1s^2 2s \,^2S_{1/2}$}  			&	\multicolumn{2}{c|}{$\to1s^2 2p \,^2P_{1/2}$}  			&	\multicolumn{2}{c|}{$\to1s^2 2p \,^2P_{3/2}$}  	\\
\multicolumn{1}{c|}{Max. $n$}  	&	\multicolumn{1}{c}{Ener.}  	&	\multicolumn{1}{c|}{Width}  	&	\multicolumn{1}{c}{Ener.}  	&	\multicolumn{1}{c|}{Width}  	&	\multicolumn{1}{c}{Ener.}  	&	\multicolumn{1}{c|}{width}  	&	\multicolumn{1}{c}{Ener.}  	&	\multicolumn{1}{c|}{Width}  	&	\multicolumn{1}{c}{Total width}  	\\
					\hline
	\multicolumn{1}{c|}{DF}  	&	3096.57	&	62.79	&	2240.96	&	0.52	&	2208.96	&	14.36	&	2205.80	&	48.87	&	126.54	\\
2	&	3091.98	&	64.58	&	2237.06	&	24.34	&	2205.22	&	3.64	&	2201.85	&	8.83	&	101.39	\\
3	&	3091.85	&	63.43	&	2236.33	&	1.29	&	2204.44	&	2.24	&	2201.23	&	6.30	&	73.26	\\
4	&	3091.71	&	63.11	&	2236.12	&	0.22	&	2204.24	&	16.13	&	2201.06	&	49.29	&	128.75	\\
5	&	3091.68	&	63.12	&	2235.99	&	0.29	&	2204.14	&	2.34	&		&	NC	&		\\
		\end{tabular}
\end{ruledtabular}
\end{table*}																			
\endgroup

\begingroup
\squeezetable							
\begin{table*}							
\caption{							
Comparison between theoretical partial radiative widths,  Auger widths and  energies for transitions originating from the Be-like  $1s 2s^2 2p\, ^1P_1$ level. Transition energies are in \si{\eV} and  widths in  \si{\meV}. 	
}
\label{tab:be-rate-theory} 							
			%
\begin{ruledtabular}
\begin{tabular}{cc|dd|dd|dd}	
	&		&	\multicolumn{2}{c|}{MCDF, Chen (1985) \cite{che1985}}			&	\multicolumn{2}{c|}{MCDF, Costa \etal 2001 \cite{cmps2001}}			&	\multicolumn{2}{c}{RCI, Natarajan (2003) \cite{nat2003}}			\\
Initial Level	&	final level	&	\multicolumn{1}{c}{energy}	&	\multicolumn{1}{c|}{rate}	&	\multicolumn{1}{c}{energy}	&	\multicolumn{1}{c|}{rate}	&	\multicolumn{1}{c}{energy}	&	\multicolumn{1}{c}{rate}	\\
$1s 2s^2 2p\, ^1P_1$	&	$1s^2 2s^2 \, ^1S_0$	&	3090.66	&	66.48	&	3091.95	&	64.57	&	3088.958	&	64.58	\\
	&		&		&		&		&		&		&		\\
$1s 2s^2 2p\, ^1P_1$	&	$1s^2 2s_{1/2}$	&	2236.81	&		&	2237.03	&	18.76	&		&		\\
	&	$1s^2 2p_{1/2}$	&	2204.79	&		&	2205.19	&	15.01	&		&		\\
	&	$1s^2 2p_{3/2}$	&	2201.63	&		&	2201.82	&	52.53	&		&		\\
Total Auger	&		&		&	80.30	&		&	86.29	&		&		\\
Level width	&		&		&	146.78	&		&	150.86	&		&		\\
\cline{1-8}
	&		&	\multicolumn{2}{c|}{MCDF (this work)}			&	\multicolumn{2}{c|}{FAC (this work)}			&		\multicolumn{2}{c}{}			\\
	&		&	\multicolumn{1}{c}{energy}	&	\multicolumn{1}{c|}{rate}	&	\multicolumn{1}{c}{energy}	&	\multicolumn{1}{c|}{rate}	&		&	\multicolumn{1}{c}{}	\\
$1s 2s^2 2p\, ^1P_1$	&	$1s^2 2s^2 \, ^1S_0$	&	3091.72	&	63.12	&	3091.11	&	63.48	&		&		\\
	&		&		&		&		&		&		&		\\
$1s 2s^2 2p\, ^1P_1$	&	$1s^2 2s_{1/2}$	&	2235.99	&	0.29	&	2241.39	&	1.13	&		&		\\
	&	$1s^2 2p_{1/2}$	&	2204.14	&	2.34	&	2209.22	&	12.93	&		&		\\
	&	$1s^2 2p_{3/2}$	&	2201.06	&	49.31	&	2206.10	&	43.82	&		&		\\
Total Auger	&		&		&	51.94	&		&	57.89	&		&		\\
Level width	&		&		&	128 (40)	&		&	121.36	&		&		\\
\end{tabular}
\end{ruledtabular}
\end{table*}																			
\endgroup

\section{Results and comparison with theory for the He-like $1s2p \,^1P_1 \rightarrow 1s 2s\,^1S_0$ transition}
\label{sec:results-he}

\subsection{Line widths}
\label{subsec:res_width_he}

Our experimental values for the line widths, obtained as explained in Sec. \ref{subsec:widths} and   Fig. \ref{fig:whe-like}, are presented in Table \ref{tab:lifetimes-he}, together with several theoretical results.  There are several possible E1 radiative  transitions originating from the $1s2p \,^1P_1$ level. Because of the large energy difference, the contribution of the $1s2p \,^1P_1 \rightarrow 1s^2\,^1S_0$ transition to the level width is strongly dominant. The next largest contribution, due to the $1s2p \,^1P_1 \rightarrow 1s 2s\,^1S_0$ transition, contributes only \SI{0.0001}{\meV} to the \SI{70.4}{\meV} width. The width of the $n=2\to n=1$ transitions has been calculated using Drake's unified method \cite{dra1979}, relativistic random phase approximation, MCDF, relativistic configuration interaction (RCI) and QED \cite{jps1995}. The effect of the negative energy continuum has been discussed in Refs. \cite{ind1996,dsjp1998}. Radiative corrections to the photon emission have also been evaluated \cite{isv2004}. The differences between all theoretical values and our measurement are well within the experimental error bar.

%
%
\begin{table}							
\caption{							
Measured and computed natural line width values for the $1s2p \,^1P_1 \rightarrow 1s^2\,^1S_0$ transitions in He-like Ar. All values are given in \si{\meV}, and estimated uncertainties are shown in parentheses. 				
}
\label{tab:lifetimes-he} 							
\begin{ruledtabular}
\begin{tabular}{clll}							
 \multicolumn{1}{c}{Transition}&  \multicolumn{1}{c}{Experiment }	& \multicolumn{1}{c}{Theory } & Reference\\
 \hline
$1s2p \,^1P_1 \rightarrow1s^2 \,^1S_0$ 			& 75 (17)	& 70.4778 (25) & MCDF (this work)\\
 & & 70.40 & MBPT, Si \etal (2016) \cite{sgwl2016} \\
 & & 70.43 & MCDHF, Si \etal (2016) \cite{sgwl2016} \\
 & & 70.43 & Johnson \etal (1995) \cite{jps1995} \\
 & & 70.49 (14)          & Drake (1979) \cite{dra1979} \\
\end{tabular}							
\end{ruledtabular}
\end{table}

\subsection{Transition energies}
\label{subsec:res_ener}

We present in Fig. \ref{fig:he1p1ener}  the transition energy values obtained from the successive pairs of dispersive and nondispersive-modes spectra, recorded during the experiment for the He-like argon $1s2p \,^1P_1 \rightarrow1s^2 \,^1S_0$  following the method presented in Sec. \ref{sec:data-analysis}. The weighted average and $\pm 1 \sigma$ bands are plotted as well.

\begin{figure}
    \centering
        \includegraphics[width=\textwidth]{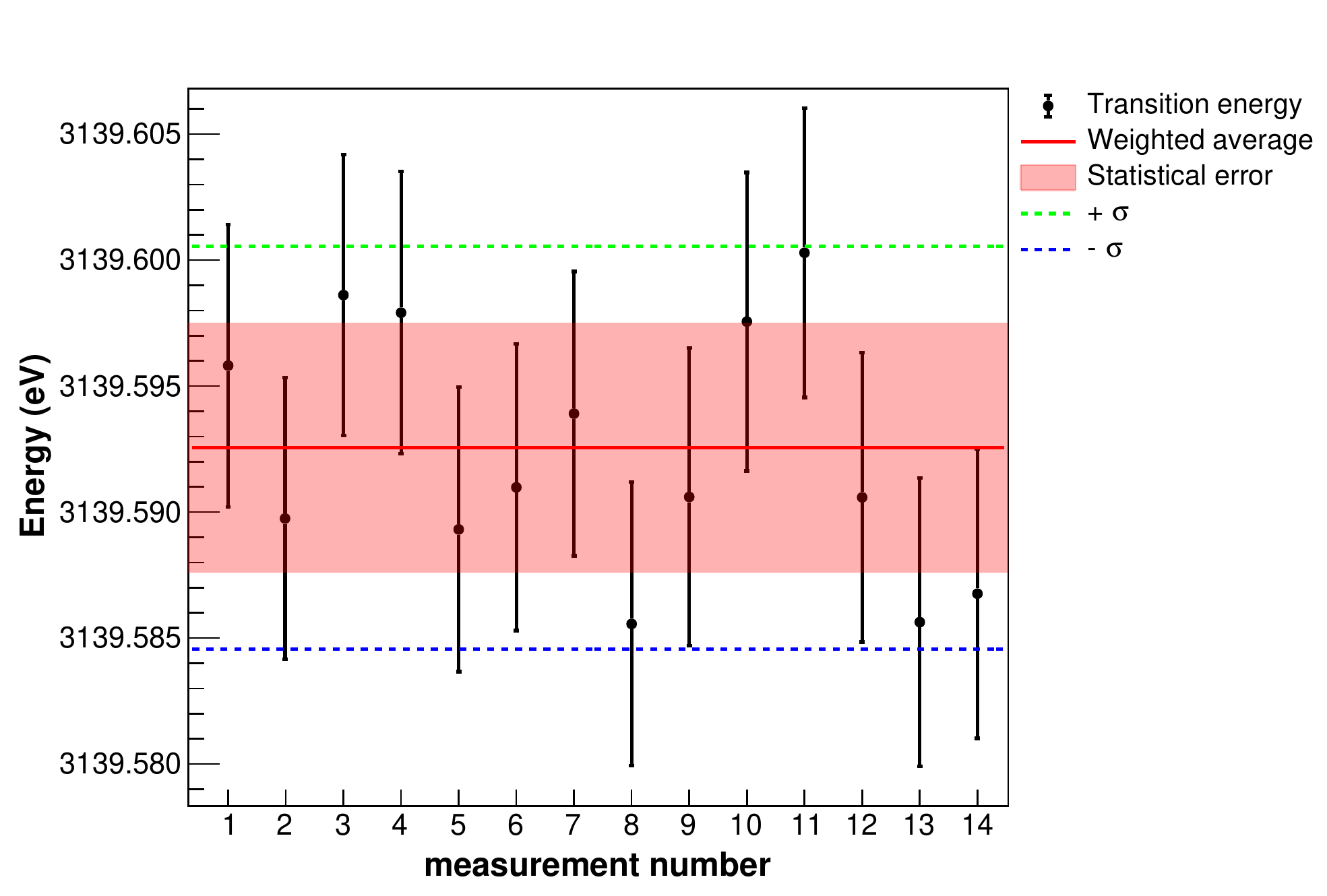}
    \caption{(Color online) He-like argon $1s2p \,^1P_1 \rightarrow 1s^2\,^1S_0$ transition energy values of the different spectra recorded during the experiment. Error bars in each point correspond to the quadratic sum of the peak fitting uncertainty with the uncertainties from Table \ref{tab:errors}, which have random fluctuations only, \ie the angle measurement and the temperature correction. The (pink) shaded area correspond to the weighted average of the peak position statistical uncertainty obtained from the fit. The $\pm  1 \sigma$ lines combine this statistical uncertainties with all systematic errors from Table \ref{tab:errors}.
    Every pair of points correspond to one-day data taking (see text for explanations).}
        \label{fig:he1p1ener}
\end{figure}

Table~\ref{tab:he-like-result} presents the measured He-like argon $1s2p \,^1P_1 \rightarrow 1s^2\,^1S_0$ transition energy, together with all known experimental and theoretical results. The final experimental accuracy, combining the instrumental contributions from Table \ref{tab:errors} is \num{2.5E-6}. The value is in  agreement with a preliminary result, obtained with the same set-up, but using fit with Voigt profiles of both the experimental spectra and the simulations \cite{sags2013,sags2013a}.  The agreement with the most precise experiments, \textit{i.e.}, the two reference-free experiments \cite{bbkc2007,kbbl2012} and the recoil ion experiment of Deslattes \etal \cite{dbf1984} is well within combined error bars. 
The agreement with the calculation of Artemyev \etal \cite{asyp2005} is also  within the linearly combined error bars.
%
%
%

%
\begin{table}							
\caption{							
Comparison of our He-like argon experimental $1s2p \,^1P_1 \rightarrow 1s^2\,^1S_0$ transition energy 							
 with previous experimental and theoretical values.							
All energies are given in eV, and estimated uncertainties are shown in parentheses. 							
}	
\label{tab:he-like-result}							
\begin{ruledtabular}
\begin{tabular}{lll}							
Energy				&	Reference	 & Exp. method	\\
\multicolumn{2}{c}{Experiment}							\\
\cline{1-3}
3139.5927	(50)(63)(80) &	This Work	 (stat.)(syst.)(tot.)	& ECRIS \\
3139.567	(11)	&	Schlesser \etal (2013)	 ~\cite{sbcs2013}	& ECRIS \\
3139.581	(5) &	Kubi\v{c}ek \etal (2012)	 ~\cite{kbbl2012} & EBIT	\\
3139.583	(63) &	Bruhns \etal (2007)	 ~\cite{bbkc2007}	& EBIT \\
3139.552	(	37	)	&	Deslattes \etal (1984)	 ~\cite{dbf1984}	 & Recoil ions \\
3139.60	(	25	)	&	Briand \etal (1983)	 ~\cite{bmic1983} & Beam-foil 	\\
3140.1	(	7	)	&	Dohmann \etal (1979)	 ~\cite{dam1979} & Beam-foil 	\\
3138.9	(	9	)	&	Neupert \etal (1971)	 ~\cite{neu1971}& Solar emission	\\
				&			\\
\multicolumn{2}{c}{Theory}							\\
\cline{1-3}
				&			\\
3139.559 (10) (13)   & This work using model operators  \cite{sty2013,sty2015}& \\
                                 &(correlation)(SE screening) &\\
3139.553 (10) (18)     & This work using Welton model (correlation)(SE screening)&\\
3139.538           &  MBPT, Si \etal (2016) ~\cite{sgwl2016} & \\
3139.449           &  MCDHF, Si \etal (2016) ~\cite{sgwl2016} & \\
3139.5821 (4)			&	Artemyev \etal (2005)	 ~\cite{asyp2005}	&\\
3139.582				&	Plante \etal (1994)	 ~\cite{pjs1994}&	\\
3139.617				&	Cheng \etal (1994)	 ~\cite{ccjs1994}	&\\
3139.576				&	Drake (1988)	 ~\cite{dra1988}	&\\
3139.649				&	Indelicato \etal (1987)	 ~\cite{igd1987}&	\\
3139.56				&	Safronova (1981)	 ~\cite{saf1981}&	\\
3140.15				&	Johnson \etal (1976)	 ~\cite{jas1992}	&\\
3140.46				&	Gabriel (1972)	 ~\cite{gab1972}	&\\
\end{tabular}							
\end{ruledtabular}
\end{table}							

\begin{figure*}
\centering
    \begin{subfigure}[b]{0.8\textwidth}
        \includegraphics[width=\textwidth]{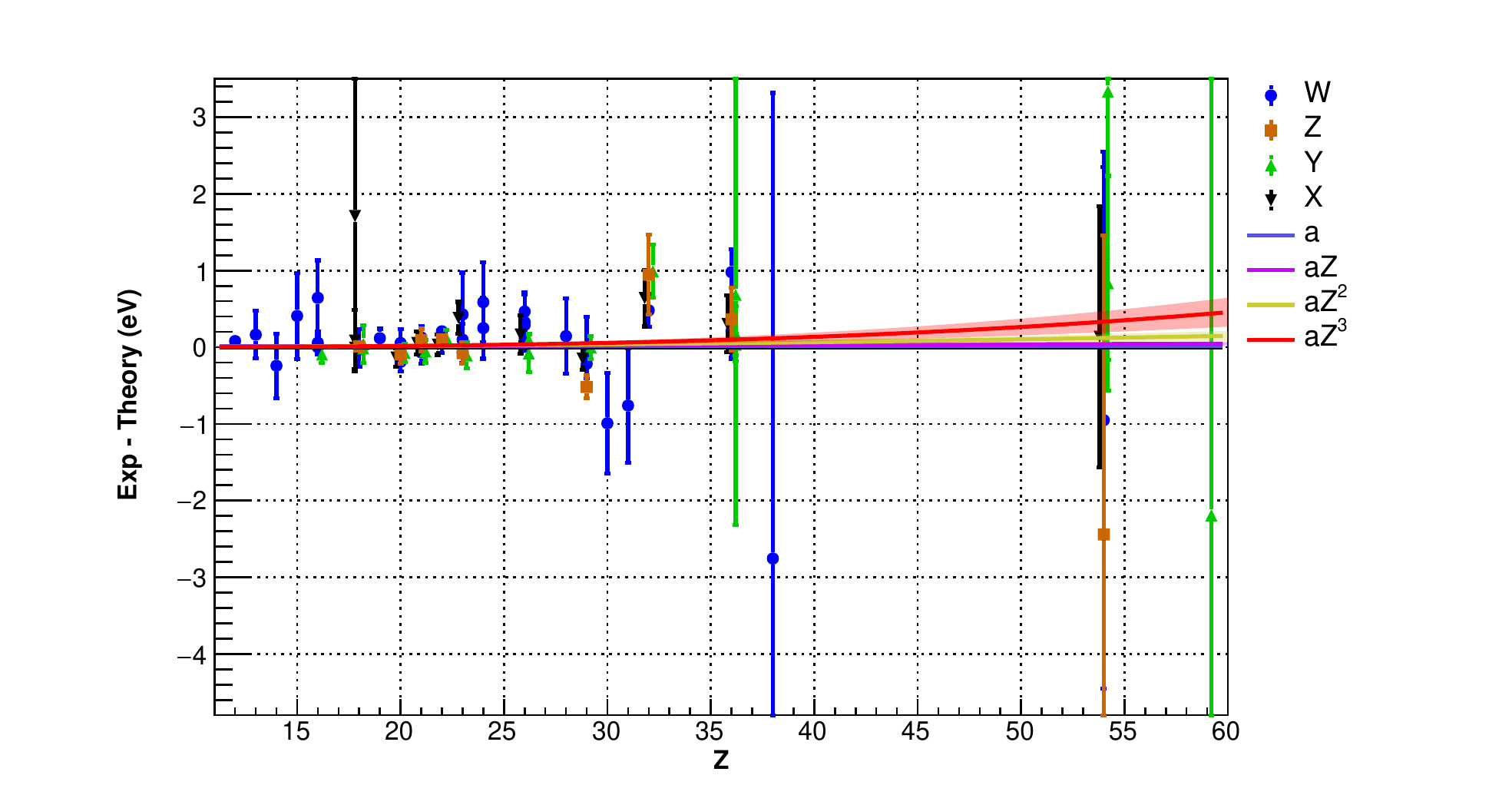}
        \caption{ $12\leq Z \leq$-range.}
        \label{fig:comp-the-exp-he-all}
    \end{subfigure}
\centering
    \begin{subfigure}[b]{0.8 \textwidth}
        \includegraphics[width=\textwidth]{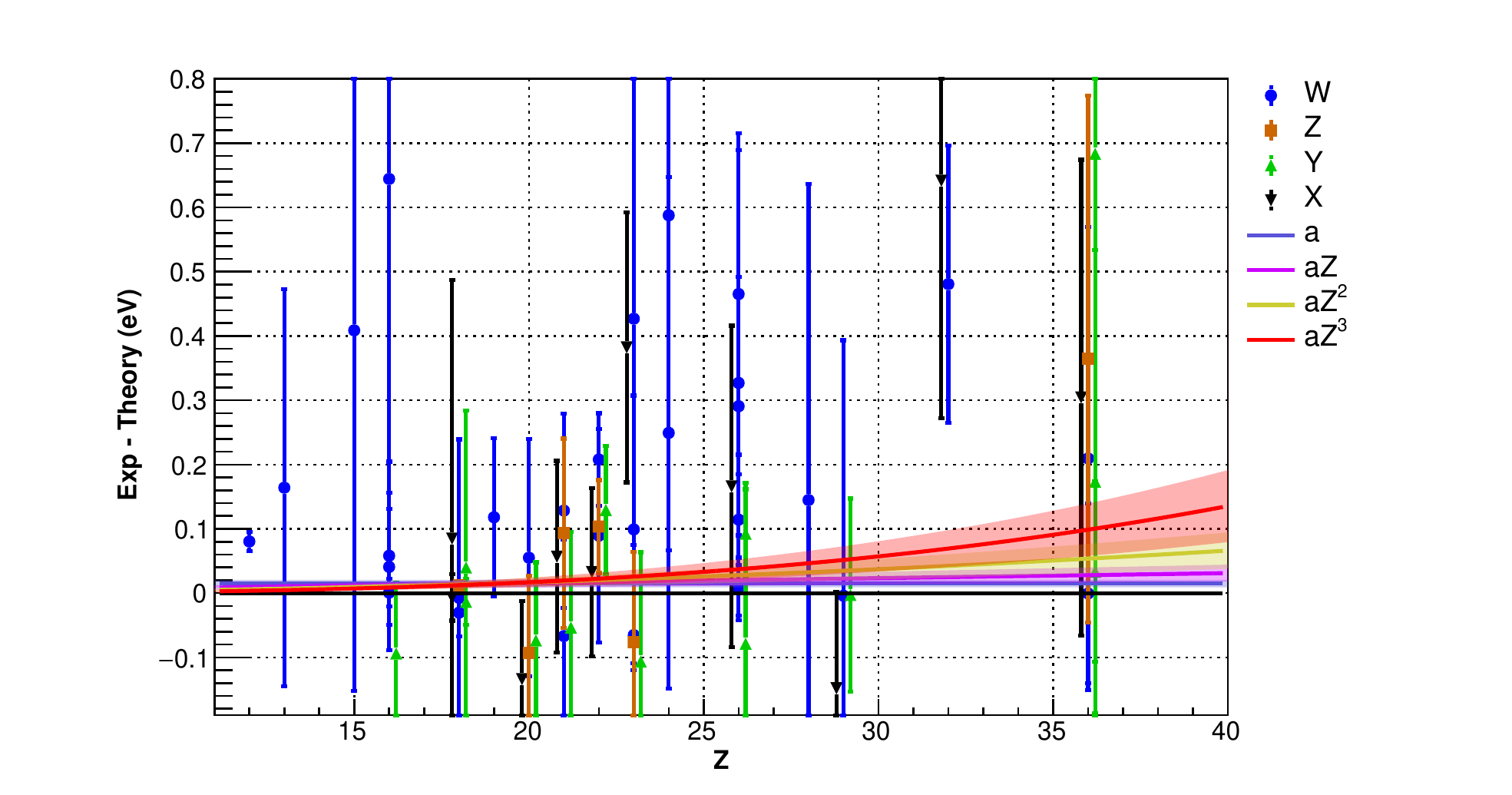}
        \caption{(Color online) Zoom on the $12\leq Z \leq 40$-range, and small energy differences.}
        \label{fig:comp-the-exp-he-zoom}
    \end{subfigure}
\caption{
(Color online) Comparison between the theoretical values by Artemyev \etal~\cite{asyp2005} and  experimental data for $n=2 \to n=1$ transition in He-like ions presented in Tables \ref{tab:he-summary} and \ref{tab:he-summary2}  for all $12\leq Z \leq 59$.
The continuous lines represent the weighted fits with  $a$, $a Z$, $a Z^2$ and $a Z^3$ functions,  and the shaded area the $\pm 1 \sigma$ bands, representing the \SI{68}{\percent} confidence interval from the fit. The experimental values for $Z=92$ are not plotted as they have very large error bars, but were included in the fit. Values of different experiments for a given $Z$ are slightly shifted horizontally to make the figure easier to read.
\label{fig:comp-the-exp-he}
}
\end{figure*}




%
\subsection{Comparison between measurements and calculations for $12\leq Z \leq 92$}
\label{sec:he_like}

There have been many measurements of $n=2\to n=1$ transition energies in He-like ions. The reference-free measurements, of the kind reported in the present work,  and the measurements calibrated against x-ray standards or transitions in H-like ions are summarized  in Tables \ref{tab:he-summary} and \ref{tab:he-summary2} for $7\leq Z \leq 92$ . Relative measurements, using the theoretical value for  one of the He-like lines in the spectrum, originating from ECRIS or Tokamak experiments are summarized in Table \ref{tab:he-relat-summary}. When older calculations were used as a reference, we used the energies of Ref. \cite{asyp2005}  to obtain an updated value for this table.  

A detailed analysis of the difference between theory  \cite{asyp2005} and experiment has been performed in previous work \cite{ckgh2012,cpgh2014,bab2015}. Here we provide an updated analysis, which include our new result and the data from Tables \ref{tab:he-summary} and \ref{tab:he-summary2} .

The differences between these experimental values and Artemyev \etal \cite{asyp2005} theoretical values are plotted in Fig. \ref{fig:comp-the-exp-he} together with weighted fits by  several functions of the shape $a Z^n$, $n=0$ to 3.  The $\pm 1\sigma$ error bands for the fits are also plotted. These error bands show that there is no significant deviation between theory and experiment.


In order to reinforce this conclusion, we have performed a systematic significance analysis. This analysis has been performed fitting functions of the form  $f(Z)=aZ^n$, $n=0$, \num{12} on three datasets build using the data presented in Tables  \ref{tab:he-summary} and \ref{tab:he-summary2}. One dataset contains only the $w$ transition, one contains all $w$, $x$, $y$, and $z$ transitions, and the last one is the same, from which the experimental values of this work, of Kubi\c{c}ek \etal \cite{kmmu2014} and of Amaro \etal  \cite{asgl2012} have been removed. The values of the reduced $\chi^2$ are plotted as a function of $n$ in Fig. \ref{fig:chi2-he} for the three subsets. It should be noted that the  reduced $\chi^2$  increases as a function of $n$, although in two of the subsets there is a weak local minimum near $n=4$. We present in Fig.~\ref{fig:signif-he} the uncertainty of the fit coefficient $a$ in standard-error units as a function of $n$ for all three datasets. The figure shows that the maximum deviation from zero is obtained for $n=0$. The deviation of the fit coefficient tends to zero with  increasing value of $n$ while the reduced $\chi^2$ increases. For the other two datasets considered, \ie all experimental values presented in Tables~\ref{tab:he-summary} and \ref{tab:he-summary2} or  the subset consisting only of the $w$-lines, there is a local maximum for each dataset around $n=4$. For all experimental data the local maximum happens at $n \simeq 4.2$ with a coefficient significance of \num{3.5} standard errors, while for the $w$-lines the local maximum is at $n \simeq 3.8$ with a deviation of \num{3} standard errors from zero. In spite of the presence of this local maximum for different monomial orders of $n$, the maximum deviation from zero of the fit parameter is at $n=0$ as well as the minimum reduced $\chi^2$ value. This leads to the conclusion that  $f(Z)=aZ^0$ is the most probable model to describe the data when considering a power law dependence with $Z$.

To sustain this conclusion, a $\chi^2$ goodness of a fit test was performed. Fig.~\ref{fig:fp_val_he} shows the result probability (p-value) of the observed $\chi^2$ cumulative distribution function (upper tail) as a function of $n$, for the given number of degrees of freedom and the minimum $\chi^2$ value of each performed fit. This probability, that the observed $\chi_{\mathrm{Obs}}^2$  for $\nu$ degrees of freedom is larger than $\chi^2$, is given by \cite{pftv2007} 
\begin{equation}
p\left(\chi^2, \nu\right)= Q\left(\frac{\chi^2}{2}, \frac{\nu}{2}\right),\nonumber\\
\end{equation}
where $Q$ is the incomplete $\Gamma$ function. When all data from Tables~\ref{tab:he-summary} and \ref{tab:he-summary2} are included, $\nu=85-1$. It can be noticed that the highest p-value for the three considered datasets is for $n=0$, and, as before, one can see a local maximum when considering all experimental results from Tables~\ref{tab:he-summary} and \ref{tab:he-summary2} or just the $w$-lines for the same $n$ value as from  Fig.~\ref{fig:signif-he}. Considering the standard significance level of \num{0.05} to evaluate the acceptance or rejection of the null hypotheses (\ie the fact  that the data can be described by the $aZ^n$ function), and since the highest p-value is \num{1.4E-6} for the three considered datasets, the null hypotheses has a very small probability to be true, with the caveats noted in Ref. \cite{wal2016}. We also performed a t-student test, which shows that $a=0$ is the most probable value for all $n$. Therefore, we conclude that it is highly unlikely that  the experiment--theory difference has a dependence in $Z$ of the form $f(Z)=aZ^n$ for any given $n$ with $0\leq n \leq 12$.

\begin{figure*}
\centering
        \includegraphics[width=\textwidth]{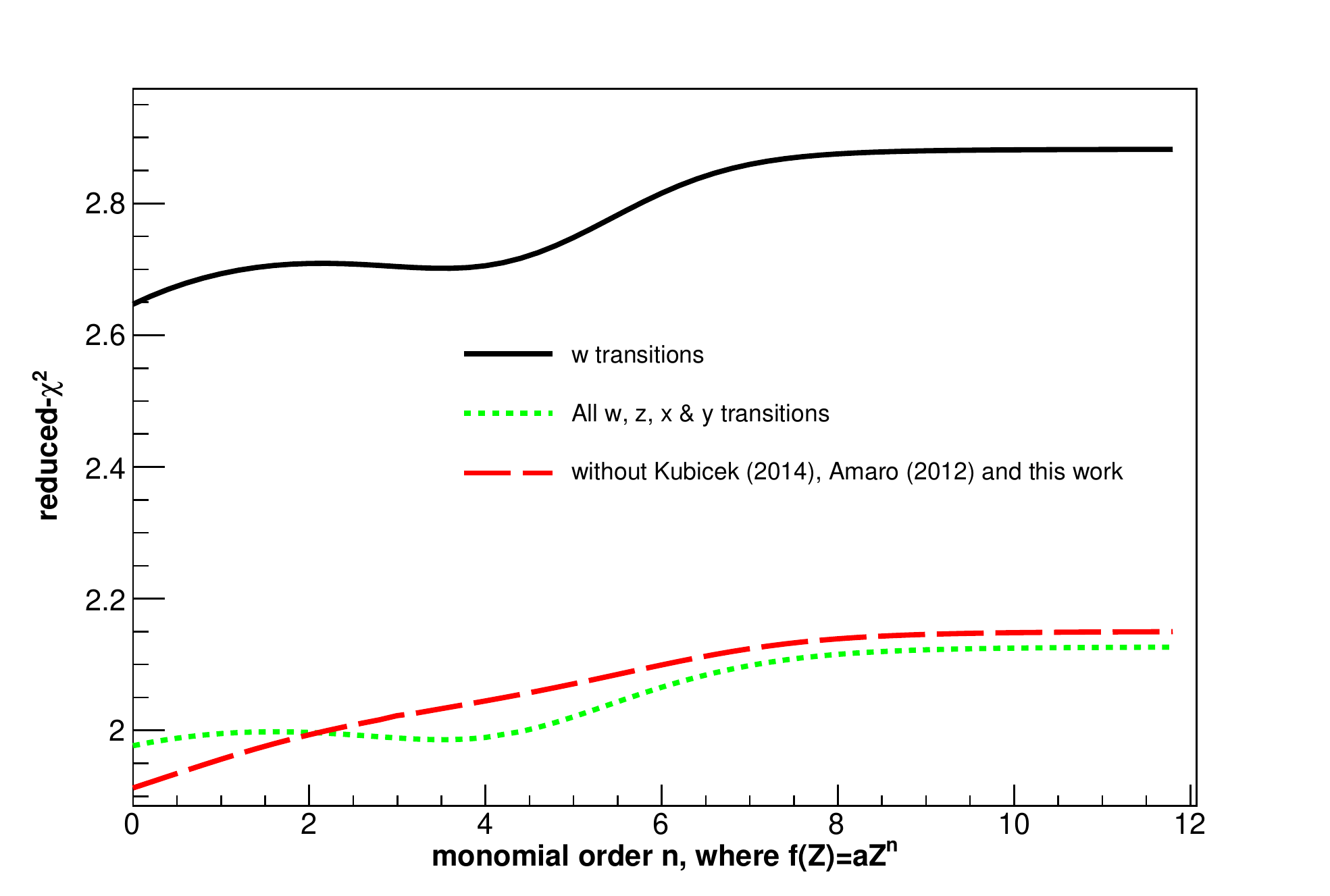}
\caption{
(Color online) Values of the reduced $\chi^2$ function as a function of $n$, when fitting $a Z^n$,  $n=0$ to 12,  to the experiment-theory differences from Tables \ref{tab:he-summary} and \ref{tab:he-summary2}. Solid line: reduced $\chi^2$ fitting only the $1s2p \,^1P_1 \rightarrow 1s^2\,^1S_0$ ($w$) values. 
Dotted line: reduced $\chi^2$ fitting all  4 $w$, $x$, $y$ and $z$ transition energies differences with theory. Dashed line: same data as dotted line, but removing the reference-free values from this work and from Refs. \cite{asgl2012,kmmu2014}.
\label{fig:chi2-he}
}
\end{figure*}

\begin{figure*}
\centering
        \includegraphics[width=\textwidth]{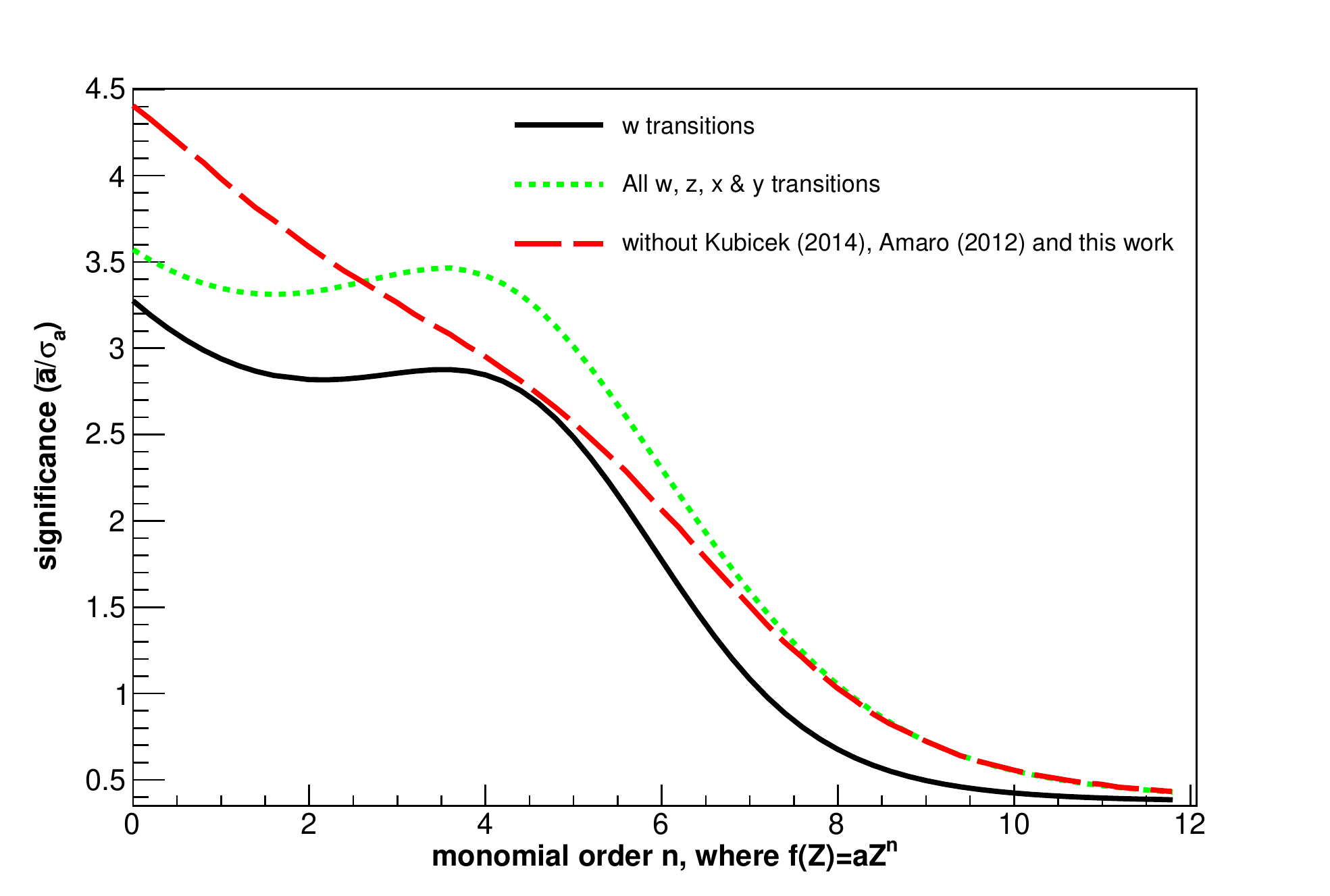}
\caption{
(Color online) Values of the significance of the fit coefficient in standard-error units as a function of $n$ when fitting $a Z^n$ to the experiment-theory differences from Tables \ref{tab:he-summary} and \ref{tab:he-summary2}.
\label{fig:signif-he}
}
\end{figure*}

\begin{figure*}
\centering
\includegraphics[width=\textwidth]{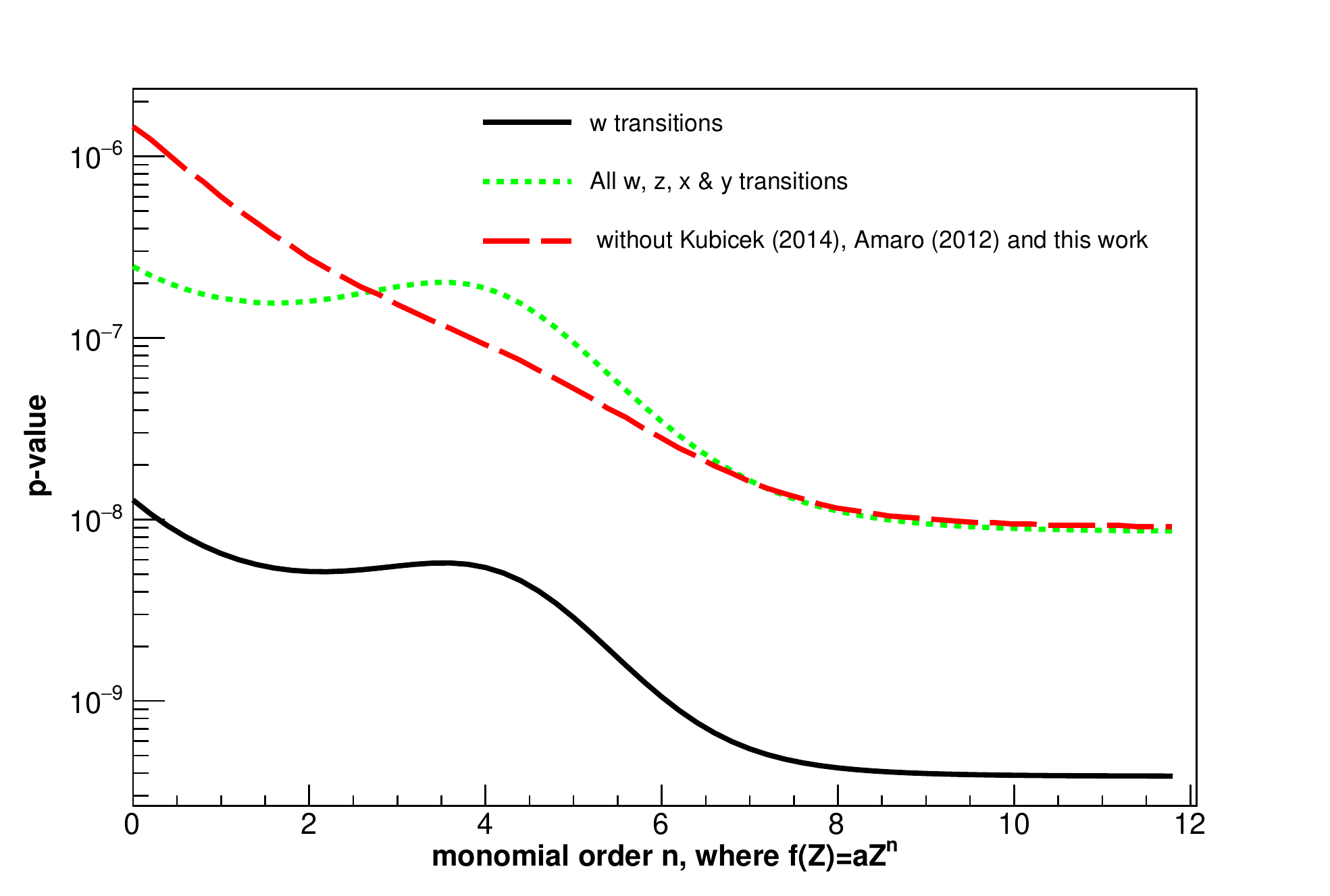}
\caption{
(Color online) $p$-value as a function of $n$ when fitting $a Z^n$ to the experiment-theory differences from Tables \ref{tab:he-summary} and \ref{tab:he-summary2}. See legend of Fig. \ref{fig:signif-he} for explanations of the data included in each curve.
\label{fig:fp_val_he}
}
\end{figure*}

\begingroup
\begin{turnpage}
\squeezetable							
\begin{table*}							
\caption{							
Summary of all measured  $n=2 \to n=1$  transition energies in He-like ions $7\leq Z \leq 20$. The theoretical values are from Ref. \cite{asyp2005}, which are available for $Z\geq12$. The experimental values are either reference-free measurements (RF) or measurements calibrated against standard reference x-ray transitions, or hydrogen-like transitions (SR). 	
}
\label{tab:he-summary} 
\begin{ruledtabular}
\begin{tabular}{dddddddddddddcc}							
	&	\multicolumn{3}{c}{ $1s2p\,^1P_1 \to 1s^2\, ^1S_0$ (w)}					&	\multicolumn{3}{c}{ $1s2p\,^3P_2 \to 1s^2\, ^1S_0$ (x)}					& 	\multicolumn{3}{c}{ $1s2p\,^3P_1 \to 1s^2\, ^1S_0$ (y)}					& 	\multicolumn{3}{c}{ $1s2s\,^3S_1 \to 1s^2\, ^1S_0$ (z)}					&		\\
$Z$	&	\multicolumn{1}{c}{Exp. (eV)}	&	\multicolumn{1}{c}{Err.}	& 	\multicolumn{1}{c}{Theory}	&	\multicolumn{1}{c}{Exp. (eV)}	&	\multicolumn{1}{c}{Err.}	& 	\multicolumn{1}{c}{Theory}	& 	\multicolumn{1}{c}{Exp. (eV)}	&	\multicolumn{1}{c}{Err.}	& 	\multicolumn{1}{c}{Theory}	& 	\multicolumn{1}{c}{Exp. (eV)}	&	\multicolumn{1}{c}{Err.}	& 	\multicolumn{1}{c}{Theory}	& \multicolumn{1}{c}{Method}	& 	\multicolumn{1}{c}{Ref.}	\\
\hline																											
7	&	430.6870	&	0.0030	& 		&		&		& 		& 		&		& 		& 		&		& 		& 	SR	& 	\cite{eal1995}	\\
8	&	573.949	&	0.011	& 		&		&		& 		& 		&		& 		& 		&		& 		& 	SR	& 	\cite{eal1995}	\\
11	&	1126.72	&	0.31	& 		&		&		& 		& 		&		& 		& 		&		& 		& 	SR	& 	\cite{abzp1974}	\\
12	&	1352.329	&	0.015	& 	1352.2483	&		&		& 	1343.5417	& 		&		& 	1343.0988	& 		&		& 	1331.1118	& 	SR	& 	\cite{abzp1974}	\\
13	&	1598.46	&	0.31	& 	1598.2914	&		&		& 	1588.7611	& 		&		& 	1588.1254	& 		&		& 	1574.9799	& 	SR	& 	\cite{abzp1974}	\\
14	&	1864.76	&	0.42	& 	1865.0014	&		&		& 	1854.6679	& 		&		& 	1853.7804	& 		&		& 	1839.4495	& 	SR	& 	\cite{abzp1974}	\\
15	&	2152.84	&	0.56	& 	2152.4310	&		&		& 	2141.3188	& 		&		& 	2140.1082	& 		&		& 	2124.5619	& 	SR	& 	\cite{abzp1974}	\\
16	&	2461.27	&	0.49	& 	2460.6292	&		&		& 	2448.7628	& 		&		& 	2447.1439	& 		&		& 	2430.3512	& 	SR	& 	\cite{abzp1974}	\\
16	&	2460.69	&	0.15	& 	2460.6292	&		&		& 	2448.7628	& 		&		& 	2447.1439	& 		&		& 	2430.3512	& 	SR	& 	\cite{aamp1988}	\\
16	&	2460.630	&	0.021	& 	2460.6292	&		&		& 	2448.7628	& 		&		& 	2447.1439	& 		&		& 	2430.3512	& 	RF	& 	\cite{kmmu2014}	\\
16	&	2460.670	&	0.090	& 	2460.6292	&		&		& 	2448.7628	& 	2447.05	&	0.11	& 	2447.1439	& 		&		& 	2430.3512	& 	SR	& 	\cite{sbbt1982}	\\
18	&		&		& 	3139.5821	&		&		& 	3126.2896	& 		&		& 	3123.5344	& 	3104.1605	&	0.0077	& 	3104.1483	& 	RF	& 	\cite{asgl2012}	\\
18	&		&		& 	3139.5821	&	3128	&	2	& 	3126.2896	& 		&		& 	3123.5344	& 		&		& 	3104.1483	& 	SR	& 	\cite{dlp1978}	\\
18	&	3139.5927	&	0.0076	& 	3139.5821	&		&		& 	3126.2896	& 		&		& 	3123.5344	& 		&		& 	3104.1483	& 	RF	& 	this work	\\
18	&	3139.5810	&	0.0092	& 	3139.5821	&		&		& 	3126.2896	& 		&		& 	3123.5344	& 		&		& 	3104.1483	& 	RF	& 	\cite{kmmu2014}	\\
18	&	3139.552	&	0.037	& 	3139.5821	&	3126.283	&	0.036	& 	3126.2896	& 	3123.521	&	0.036	& 	3123.5344	& 		&		& 	3104.1483	& 	SR	& 	\cite{dbf1984}	\\
18	&	3139.57	&	0.25	& 	3139.5821	&	3126.37	&	0.40	& 	3126.2896	& 	3123.57	&	0.24	& 	3123.5344	& 		&		& 	3104.1483	& 	SR	& 	\cite{bmic1983}	\\
19	&	3510.58	&	0.12	& 	3510.4616	&		&		& 	3496.4937	& 		&		& 	3492.9736	& 		&		& 	3472.2417	& 	SR	& 	\cite{bbvh1989}	\\
20	&	3902.43	&	0.18	& 	3902.3777	&		&		& 	3887.7607	& 		&		& 	3883.3169	& 		&		& 	3861.2059	& 	SR	& 	\cite{aamp1988}	\\
20	&	3902.19	&	0.12	& 	3902.3777	&	3887.63	&	0.12	& 	3887.7607	& 	3883.24	&	0.12	& 	3883.3169	& 	3861.11	&	0.12	& 	3861.2059	& 	SR	& 	\cite{rrag2014}	\\
21	&	4315.54	&	0.15	& 	4315.4124	&		&		& 	4300.1720	& 		&		& 	4294.6220	& 		&		& 	4271.0997	& 	SR	& 	\cite{bbvh1989}	\\
21	&	4315.35	&	0.15	& 	4315.4124	&	4300.23	&	0.15	& 	4300.1720	& 	4294.57	&	0.15	& 	4294.6220	& 	4271.19	&	0.15	& 	4271.0997	& 	SR	& 	\cite{rgtm1995}	\\
22	&	4749.73	&	0.17	& 	4749.6441	&		&		& 	4733.8008	& 		&		& 	4726.9373	& 		&		& 	4701.9746	& 	SR	& 	\cite{bbvh1989}	\\
22	&	4749.852	&	0.072	& 	4749.6441	&	4733.83	&	0.13	& 	4733.8008	& 	4727.07	&	0.10	& 	4726.9373	& 	4702.078	&	0.072	& 	4701.9746	& 	SR	& 	\cite{pckg2014}	\\
23	&	5205.59	&	0.55	& 	5205.1653	&		&		& 	5188.7378	& 		&		& 	5180.3264	& 		&		& 	5153.8962	& 	SR	& 	\cite{aamp1988}	\\
23	&	5205.26	&	0.21	& 	5205.1653	&		&		& 	5188.7378	& 		&		& 	5180.3264	& 		&		& 	5153.8962	& 	SR	& 	\cite{bbvh1989}	\\
23	&	5205.10	&	0.14	& 	5205.1653	&	5189.120	&	0.210	& 	5188.7378	& 	5180.22	&	0.17	& 	5180.3264	& 	5153.82	&	0.14	& 	5153.8962	& 	SR	& 	\cite{cphs2000}	\\
\end{tabular}
\end{ruledtabular}
\end{table*}
\end{turnpage}																			
\endgroup

\begingroup
\begin{turnpage}
\squeezetable							
\begin{table*}							
\caption{							
Summary of all measured $n=2 \to n=1$  transition energies in He-like ions $21\leq Z \leq 92$. The theoretical values are from Ref. \cite{asyp2005}. The experimental values are either reference-free measurements (RF) or measurements calibrated against standard reference x-ray transitions, or hydrogenlike transitions (SR). 	
}
\label{tab:he-summary2} 
\begin{ruledtabular}
\begin{tabular}{dddddddddddddcc}							
	&	\multicolumn{3}{c}{ $1s2p\,^1P_1 \to 1s^2\, ^1S_0$ (w)}					&	\multicolumn{3}{c}{ $1s2p\,^3P_2 \to 1s^2\, ^1S_0$ (x)}					& 	\multicolumn{3}{c}{ $1s2p\,^3P_1 \to 1s^2\, ^1S_0$ (y)}					& 	\multicolumn{3}{c}{ $1s2s\,^3S_1 \to 1s^2\, ^1S_0$ (z)}					&		\\
$Z$	&	\multicolumn{1}{c}{Exp. (eV)}	&	\multicolumn{1}{c}{Err.}	& 	\multicolumn{1}{c}{Theory}	&	\multicolumn{1}{c}{Exp. (eV)}	&	\multicolumn{1}{c}{Err.}	& 	\multicolumn{1}{c}{Theory}	& 	\multicolumn{1}{c}{Exp. (eV)}	&	\multicolumn{1}{c}{Err.}	& 	\multicolumn{1}{c}{Theory}	& 	\multicolumn{1}{c}{Exp. (eV)}	&	\multicolumn{1}{c}{Err.}	& 	\multicolumn{1}{c}{Theory}	& \multicolumn{1}{c}{Method}	& 	\multicolumn{1}{c}{Ref.}	\\
\hline																											
24	&	5682.66	&	0.52	& 	5682.0684	&		&		& 	5665.0715	& 		&		& 	5654.8491	& 		&		& 	5626.9276	& 	SR	& 	\cite{aamp1988}	\\
24	&	5682.32	&	0.40	& 	5682.0684	&		&		& 	5665.0715	& 		&		& 	5654.8491	& 		&		& 	5626.9276	& 	SR	& 	\cite{bbvh1989}	\\
26	&	6700.76	&	0.36	& 	6700.4347	&		&		& 	6682.3339	& 		&		& 	6667.5786	& 		&		& 	6636.6126	& 	SR	& 	\cite{aamp1988}	\\
26	&	6700.73	&	0.20	& 	6700.4347	&		&		& 	6682.3339	& 		&		& 	6667.5786	& 		&		& 	6636.6126	& 	SR	& 	\cite{bbvh1989}	\\
26	&	6700.441	&	0.049	& 	6700.4347	&		&		& 	6682.3339	& 		&		& 	6667.5786	& 		&		& 	6636.6126	& 	RF	& 	\cite{kmmu2014}	\\
26	&	6700.90	&	0.25	& 	6700.4347	&	6682.50	&	0.25	& 	6682.3339	& 	6667.50	&	0.25	& 	6667.5786	& 		&		& 	6636.6126	& 	SR	& 	\cite{btmd1984}	\\
26	&	6700.549	&	0.070	& 	6700.4347	&		&		& 	6682.3339	& 	6667.671	&	0.069	& 	6667.5786	& 		&		& 	6636.6126	& 	RF	& 	\cite{rbes2013}	\\
27	&	7245.88	&	0.64	& 	7242.1133	&		&		& 	7223.4718	& 		&		& 	7205.9299	& 		&		& 	7173.4164	& 	SR	& 	\cite{aamp1988}	\\
28	&	7805.75	&	0.49	& 	7805.6053	&		&		& 	7786.4246	& 		&		& 	7765.7048	& 		&		& 	7731.6307	& 	SR	& 	\cite{aamp1988}	\\
29	&	8391.03	&	0.40	& 	8391.0349	&		&		& 	8371.3181	& 		&		& 	8346.9929	& 		&		& 	8311.3467	& 	SR	& 	\cite{aamp1988}	\\
29	&	8390.82	&	0.15	& 	8391.0349	&	8371.17	&	0.15	& 	8371.3181	& 	8346.99	&	0.15	& 	8346.9929	& 	8310.83	&	0.15	& 	8311.3467	& 	SR	& 	\cite{bab2015}	\\
30	&	8997.53	&	0.65	& 	8998.5238	&		&		& 	8978.2677	& 		&		& 	8949.8740	& 		&		& 	8912.6466	& 	SR	& 	\cite{aamp1988}	\\
31	&	9627.45	&	0.75	& 	9628.2072	&		&		& 	9607.4099	& 		&		& 	9574.4461	& 		&		& 	9535.6292	& 	SR	& 	\cite{aamp1988}	\\
32	&	10280.70	&	0.22	& 	10280.2175	&	10259.52	&	0.37	& 	10258.8739	& 	10221.79	&	0.35	& 	10220.7996	& 	10181.33	&	0.52	& 	10180.3868	& 	SR	& 	\cite{mbvk1992}	\\
36	&	13115.45	&	0.30	& 	13114.4705	&		&		& 	13090.8657	& 	13026.8	&	3.0	& 	13026.1165	& 		&		& 	12979.2656	& 	SR	& 	\cite{itbl1986}	\\
36	&	13114.68	&	0.36	& 	13114.4705	&	13091.17	&	0.37	& 	13090.8657	& 	13026.29	&	0.36	& 	13026.1165	& 	12979.63	&	0.41	& 	12979.2656	& 	SR	& 	\cite{wbdb1996}	\\
36	&	13114.47	&	0.14	& 	13114.4705	&		&		& 	13090.8657	& 	13026.15	&	0.14	& 	13026.1165	& 		&		& 	12979.2656	& 	RF	& 	\cite{esbr2015}	\\
38	&	14666.8	&	6.1	& 	14669.5399	&		&		& 	14644.7518	& 		&		& 	14562.2995	& 		&		& 	14512.1996	& 	SR	& 	\cite{aamp1988}	\\
39	&	15475.6	&	2.9	& 	15482.1565	&		&		& 	15456.7619	& 		&		& 	15364.1984	& 		&		& 	15312.4664	& 	SR	& 	\cite{aamp1988}	\\
54	&	30629.1	&	3.5	& 	30630.0512	&		&		& 	30594.3635	& 	30209.6	&	3.5	& 	30206.2652	& 		&		& 	30129.1420	& 	SR	& 	\cite{biss1989}	\\
54	&	30619.9	&	4.0	& 	30630.0512	&		&		& 	30594.3635	& 	30210.5	&	4.5	& 	30206.2652	& 	30126.70	&	3.90	& 	30129.1420	& 	SR	& 	\cite{wbbc2000}	\\
54	&	30631.2	&	1.2	& 	30630.0512	&	30594.50	&	1.70	& 	30594.3635	& 	30207.1	&	1.4	& 	30206.2652	& 		&		& 	30129.1420	& 	SR	& 	\cite{tgbb2009}	\\
59	&		&		& 	37003.7270	&		&		& 	36964.0900	& 	36389.1	&	6.8	& 	36391.2920	& 		&		& 	36305.1570	& 	SR	& 	\cite{tbcc2008}	\\
92	&	100626	&	35	& 	100610.89	&		&		& 	100537.18	& 		&		& 	96169.63	& 		&		& 	96027.15	& 	SR	& 	\cite{bcid1990}	\\
92	&	100598	&	107	& 	100610.89	&		&		& 	100537.18	& 		&		& 	96169.63	& 		&		& 	96027.15	& 	SR	& 	\cite{ldhs1994}	\\
\end{tabular}
\end{ruledtabular}
\end{table*}
\end{turnpage}																			
\endgroup

\begingroup
\begin{turnpage}
\squeezetable							
\begin{table*}							
\caption{							
Summary of all   $n=2 \to n=1$  transition energies in He-like ions $Z\geq7$, calibrated relative to the theoretical value of one of the four He-like transition (x, y, z or w). The line used as calibration is noted ``Ref.''. The energies of the measured lines have been re-evaluated using Ref.  \cite{asyp2005} for the reference transition energy. The displayed theoretical values are also from Ref. \cite{asyp2005}.	
}
\label{tab:he-relat-summary} 
\begin{ruledtabular}
\begin{tabular}{dddddddddddddc}							
	&	\multicolumn{3}{c}{ $1s2p\,^1P_1 \to 1s^2\, ^1S_0$ (w)}					&	\multicolumn{3}{c}{ $1s2p\,^3P_2 \to 1s^2\, ^1S_0$ (x)}					& 	\multicolumn{3}{c}{ $1s2p\,^3P_1 \to 1s^2\, ^1S_0$ (y)}					& 	\multicolumn{3}{c}{ $1s2s\,^3S_1 \to 1s^2\, ^1S_0$ (z)}					&		\\
$Z$	&	\multicolumn{1}{c}{Exp. (eV)}	&	\multicolumn{1}{c}{Err.}	& 	\multicolumn{1}{c}{Theory}	&	\multicolumn{1}{c}{Exp. (eV)}	&	\multicolumn{1}{c}{Err.}	& 	\multicolumn{1}{c}{Theory}	& 	\multicolumn{1}{c}{Exp. (eV)}	&	\multicolumn{1}{c}{Err.}	& 	\multicolumn{1}{c}{Theory}	& 	\multicolumn{1}{c}{Exp. (eV)}	&	\multicolumn{1}{c}{Err.}	& 	\multicolumn{1}{c}{Theory}	& 	\multicolumn{1}{c}{Ref.}	\\
\hline																											
16	&		&		& 	2460.6292	&	2448.739	&	0.020	& 	2448.7628	& 	2447.150	&	0.009	& 	2447.1439	& 	\multicolumn{2}{c}{Ref.}			& 	2430.3512	& 	\cite{sbcs2013}	\\
18	&	3139.567	&	0.011	& 	3139.5821	&	3126.291	&	0.011	& 	3126.2896	& 	3123.489	&	0.012	& 	3123.5344	& 	\multicolumn{2}{c}{Ref.}			& 	3104.1483	& 	\cite{sbcs2013}	\\
18	&	\multicolumn{2}{c}{Ref.}			& 	3139.5821	&	3126.440	&	0.079	& 	3126.2896	& 	3123.604	&	0.079	& 	3123.5344	& 	3104.21	&	0.16	& 	3104.1483	& 	\cite{tbbf1985}	\\
21	&	\multicolumn{2}{c}{Ref.}			& 	4315.4124	&	4300.00	&	0.30	& 	4300.1720	& 	4294.49	&	0.30	& 	4294.6220	& 	4271.99	&	0.29	& 	4271.0997	& 	\cite{tcdl1985}	\\
22	&	\multicolumn{2}{c}{Ref.}			& 	4749.6441	&	4733.86	&	0.18	& 	4733.8008	& 	4726.82	&	0.18	& 	4726.9373	& 	4701.89	&	0.18	& 	4701.9746	& 	\cite{bhzv1985}	\\
23	&	\multicolumn{2}{c}{Ref.}			& 	5205.1653	&	5188.18	&	0.43	& 	5188.7378	& 	5179.51	&	0.43	& 	5180.3264	& 	5153.24	&	0.43	& 	5153.8962	& 	\cite{tcdl1985}	\\
24	&	\multicolumn{2}{c}{Ref.}			& 	5682.0684	&	5664.67	&	0.52	& 	5665.0715	& 	5654.60	&	0.52	& 	5654.8491	& 	5626.63	&	0.51	& 	5626.9276	& 	\cite{tcdl1985}	\\
25	&	\multicolumn{2}{c}{Ref.}			& 	6180.4573	&	6163.25	&	0.61	& 	6162.9043	& 	6150.11	&	0.61	& 	6150.5777	& 	6120.66	&	0.60	& 	6121.1432	& 	\cite{tcdl1985}	\\
28	&	\multicolumn{2}{c}{Ref.}			& 	7805.6053	&	7786.96	&	0.49	& 	7786.4246	& 		&		& 	7765.7048	& 		&		& 	7731.6307	& 	\cite{hbhv1987}	\\
\end{tabular}
\end{ruledtabular}
\end{table*}
\end{turnpage}																			
\endgroup

%
\section{Results and comparison with theory for the Be-like $1s 2s^2 2p \,^1P_1 \rightarrow 1s^2 2s^2\,^1S_0$ transition}
\label{sec:results-be}

\begin{figure}
\centering
\includegraphics[clip=true,width=13cm]{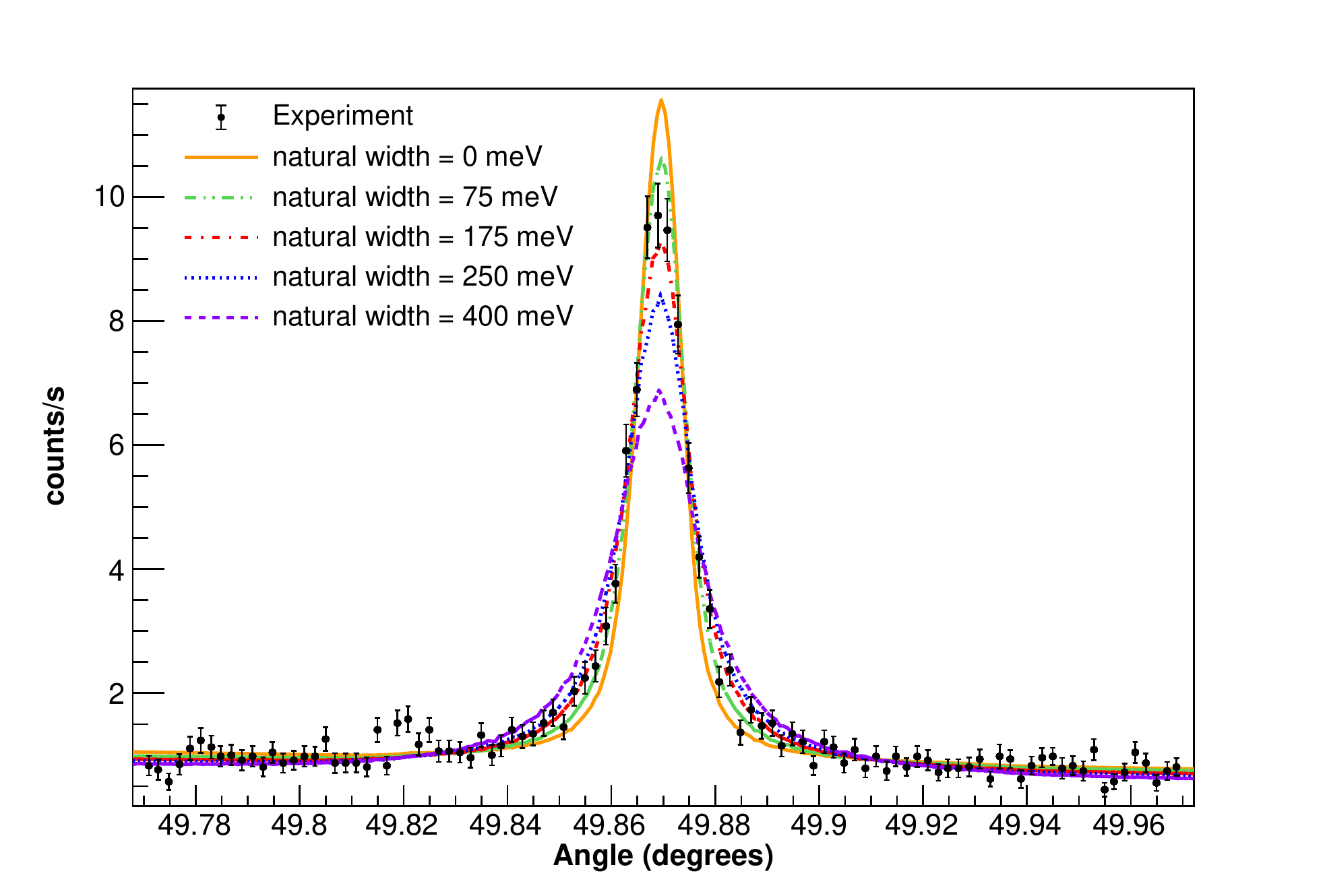}
\caption{
(Color online) Example of a dispersive-mode experimental spectrum for the Be-like Ar $1s 2s^2 2p \,^1P_1 \rightarrow 1s^2 2s^2\,^1S_0$ transition (black dots), together with a few plots of the function in Eq. \eqref{eq:fit_func},
for different values of the natural line width  $\Gamma_{\textrm{L}}^i$. The four parameters have been adjusted to minimize the reduced $\chi^2\left(\Gamma_{\textrm{L}}\right)$ (see text for more explanations).
\label{fig:be-width-ex}
}
\end{figure}

A typical spectrum  for the $1s 2s^2 2p \,^1P_1 \rightarrow 1s^2 2s^2\,^1S_0$ transition, obtained in  dispersive mode, is presented in Fig. \ref{fig:be-width-ex}. The width of the  $1s 2s^2 2p \,^1P_1 $ in contrast to the He-like case, has both radiative and non-radiative (Auger) contributions. The radiative part is also heavily dominated by the $1s 2s^2 2p \,^1P_1 \rightarrow 1s^2 2s^2\,^1S_0$ transition. As seen in Table \ref{tab:be-rate-theory}, the non-radiative part is mostly due to three Auger transitions, the  $1s 2s^2 2p \,^1P_1 \rightarrow 1s^2 2s\,^2S_{1/2}$, the  $1s 2s^2 2p \,^1P_1 \rightarrow 1s^2 2p\,^2P_{1/2}$ and the $1s 2s^2 2p \,^1P_1 \rightarrow 1s^2 2p\,^2P_{3/2}$. The radiative and non-radiative contributions are of similar size. The distribution of results from the daily experiments is presented in Fig. \ref{fig:wbe-like}. Our experimental width and the comparison with theory are presented in Table \ref {tab:lifetimes-be}. The agreement between theory and experiment is  within combined experimental and theoretical uncertainty.

We present in Fig. \ref{fig:be1p1ener}  the transition energy values obtained from the successive pairs of dispersive and nondispersive-mode spectra, recorded during the experiment for the  $1s2s^2 2p \,^1P_1 \rightarrow 1s^2 2s^2\,^1S_0$ transition, following the method presented in Sec. \ref{sec:data-analysis}. The weighted average and $\pm 1 \sigma$ values are plotted as well.

\begin{figure}
    \centering
    
    ~ 
        \includegraphics[width=\textwidth]{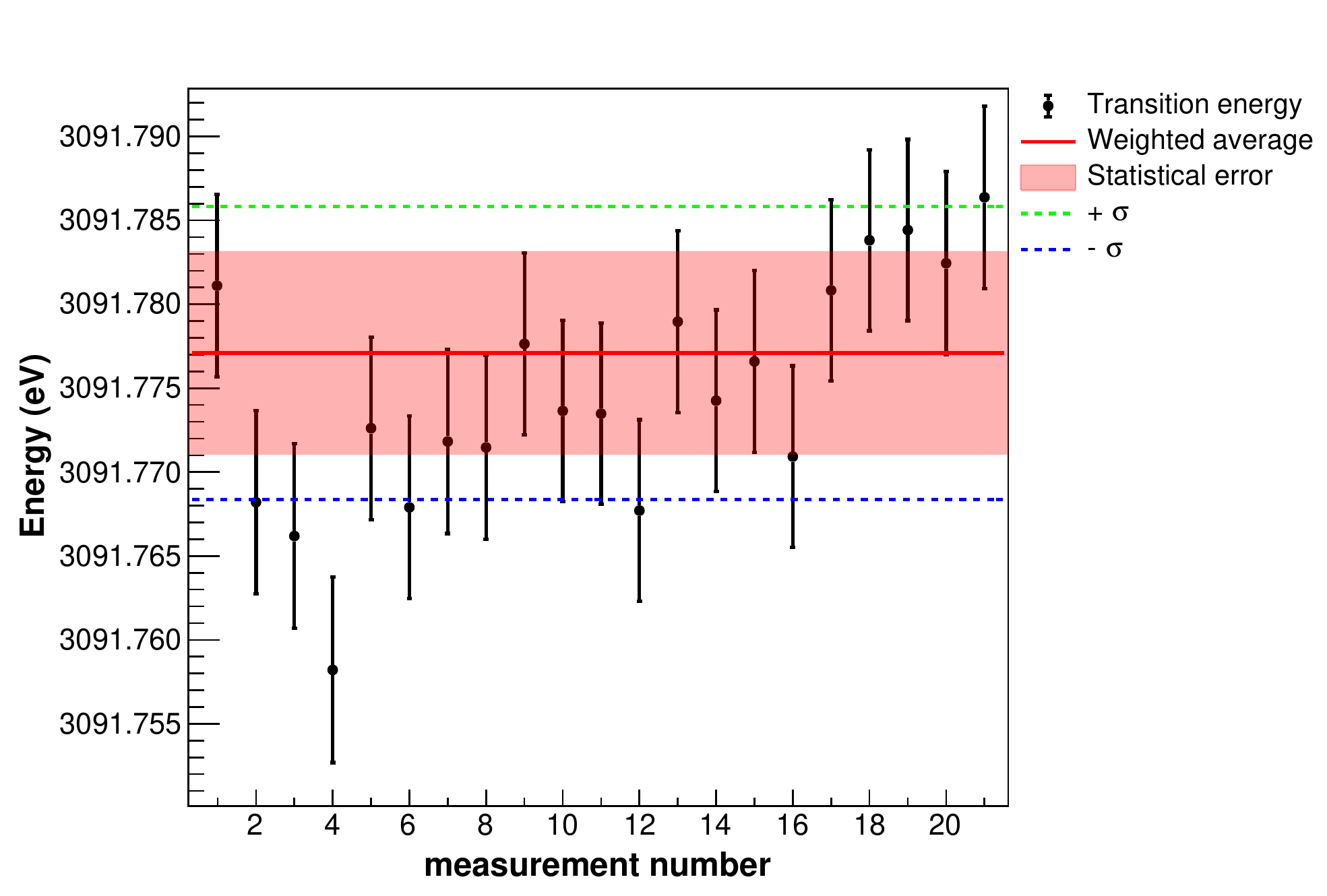}
    \caption{(Color online) Be-like argon $1s2s^22p \,^1P_1 \rightarrow 1s^22s^2\,^1S_0$ transition  energy values for the different spectra recorded during the experiment. Error bars in each point correspond to the quadratic sum of the peak fitting uncertainty with the uncertainties from Table \ref{tab:errors}, which have random fluctuations only, \ie the angle measurement and the temperature correction. The (pink) shaded area correspond to the weighted average of the peak position statistical uncertainty obtained from the fit. The $\pm  1 \sigma$ lines combine this statistical uncertainties with all systematic errors from Table \ref{tab:errors}.    Every pair of points correspond to one-day data taking (see text for explanations).}
        \label{fig:be1p1ener}
\end{figure}

\begin{table}							
\caption{							
Measured and computed natural line width values for the  $1s 2s^2 2p \,^1P_1 \rightarrow 1s^2 2s^2\,^1S_0$ transition in Be-like Ar. All values are given in \si{\meV}, and estimated uncertainties are shown in parentheses. 				
}
\label{tab:lifetimes-be} 							
\begin{ruledtabular}
\begin{tabular}{clll}							
 \multicolumn{1}{c}{Transition}&  \multicolumn{1}{c}{Experiment }	& \multicolumn{1}{c}{Theory } & Reference\\
 \hline
 $1s2s^2 2p \,^1P_1 \rightarrow 1s^2 2s^2\,^1S_0$	& 146 (18)	& 128 (40) & MCDF (this work)\\
 && 121.4& FAC (this work)\\
  & & 150.9 & Costa \etal (2001) \cite{cmps2001} \\
  & & 146.8 & Chen (1985) \cite{che1985} \\
  & & 106.1 & Safronova \etal (1979) \cite{sal1979} \\
\end{tabular}							
\end{ruledtabular}
\end{table}																			

In Table~\ref{tab:be-ener}, we present our results for the $1s2s^2 2p \,^1P_1 \rightarrow 1s^22s^2\,^1S_0$ transition energies. The measurement has been performed with a relative uncertainty of \num{2.8E-6}. The difference with Yerokhin \etal calculation  \cite{ysf2015}, which is given with a relative accuracy of \num{11E-6}, is \num{9.7E-6}. The difference with our MCDF results using effective operators self-energy screening is \num{2.3E-6}, while it is \num{3.6E-6} with the calculation using the Welton method.
The difference between the present reference-free measurement and the relative measurement presented in Ref. \cite{sbcs2013}, calibrated against the theoretical value of the $1s 2s \,^3S_1 \to 1s^2 \,^1S_0$ transition energy of \cite{asyp2005} is only \num{0.4E-6}. All recent measurements and calculations are thus forming a very coherent set of data.

The energy of this transition has not been extensively studied. It was measured relative either to  theoretical values in S, Cl and Ar \cite{sbcs2013}, Sc \cite{rgtm1995}, Fe \cite{bpjh1993,dbkj1997}, Ni \cite{hbhv1987} and Pr \cite{tbcc2008} or to K-edges in Fe \cite{rbes2013}. The  width and Auger rate for this transition have also been measured in iron \cite{rbes2013,sber2015}, with the combined use of synchrotron radiation and ion production with an EBIT.
 In Fig. \ref{fig:be-z-comp}, we present a comparison between theory and experiment, and between different calculations for the $1s2s^2 2p \,^1P_1 \rightarrow 1s^22s^2\,^1S_0$ line energy, for $10\leq Z \leq 29$. 
Since there is no recent  calculation covering all elements for which there is a measured value, we use as reference the old calculation from Ref. \cite{sal1979}, which does not include accurate QED corrections.

To conclude the discussion on both transitions measured here, we have subtracted the $1s2s\,^3S_1 \to 1s^2\, ^1S_0$ M1 transition energy measured  with the same method in Ref. \cite{asgl2012}  from the  energies of the  $1s2p\,^1P_1 \to 1s^2\, ^1S_0$ and the $1s2s^2 2p \,^1P_1 \rightarrow 1s^2 2s^2\,^1S_0$ transition energies measured here (Table \ref{tab:relat-ener}). The agreement with the relative measurements performed in Ref. \cite{sbcs2013} is within combined error bars. The difference between the reference-free transition measurements are in even better agreement with theory than the direct measurements reported in  Ref. \cite{sbcs2013}.

%
\begingroup
\squeezetable							
\begin{table}							
\caption{							
Comparison between experimental and theoretical  Be-like argon $1s2s^2 2p \,^1P_1 \rightarrow 1s^2 2s^2\,^1S_0$ transition energies. All energies are given in eV, and estimated uncertainties are shown in parentheses. 
}
\label{tab:be-ener} 							
\begin{ruledtabular}
\begin{tabular}{dc}							
\multicolumn{1}{c}{Transition energy}  &	Reference		\\ 
\cline{1-2} 
\multicolumn{2}{c}{Experiment}							\\
3091.7771(61)(63)(87) &	This work	(stat.)(syst.)(tot.)	\\
3091.776(3)& Schlesser \etal (2013) \cite{sbcs2013}   \\
\cline{1-2} 
\multicolumn{2}{c}{Theory} \\
3091.716	(30)(18)(11)				&	This work using model operators  \cite{sty2013,sty2015}	(see Table \ref{tab:be-like-total-energy}) (Corr.)(SE screening)(Auger shift)\\
3091.710 (30)(16)(11)			&	This work using Welton model	(see Table \ref{tab:be-like-total-energy})	(Corr.)(SE screening)(Auger shift)\\
3091.11 &	This work using FAC \cite{gu2008} \\ 
3091.749 		(34) &	Yerokhin \etal (2015)	 \cite{ysf2015}	\\
3088.958 & Natarajan (2003) \cite{nat2003} \\
 3091.95                & Costa \etal (2001) \cite{cmps2001} \\
3092.157 &  Safronova and Shlyaptseva (1996) \cite{sas1996} \\
3090.64   & Chen and Crasemann (1987)  \cite{cac1987} \\
3090.66   & Chen  (1985)  \cite{che1985} \\
3092.18   & Safronova and Lisina (1979) \cite{sal1979}\\
3092.18  & Boiko \etal (1978) \cite{bcif1978} \\%
\end{tabular}							
\end{ruledtabular}
\end{table}																			
\endgroup

\begin{figure*}
\centering
        \includegraphics[width=\textwidth]{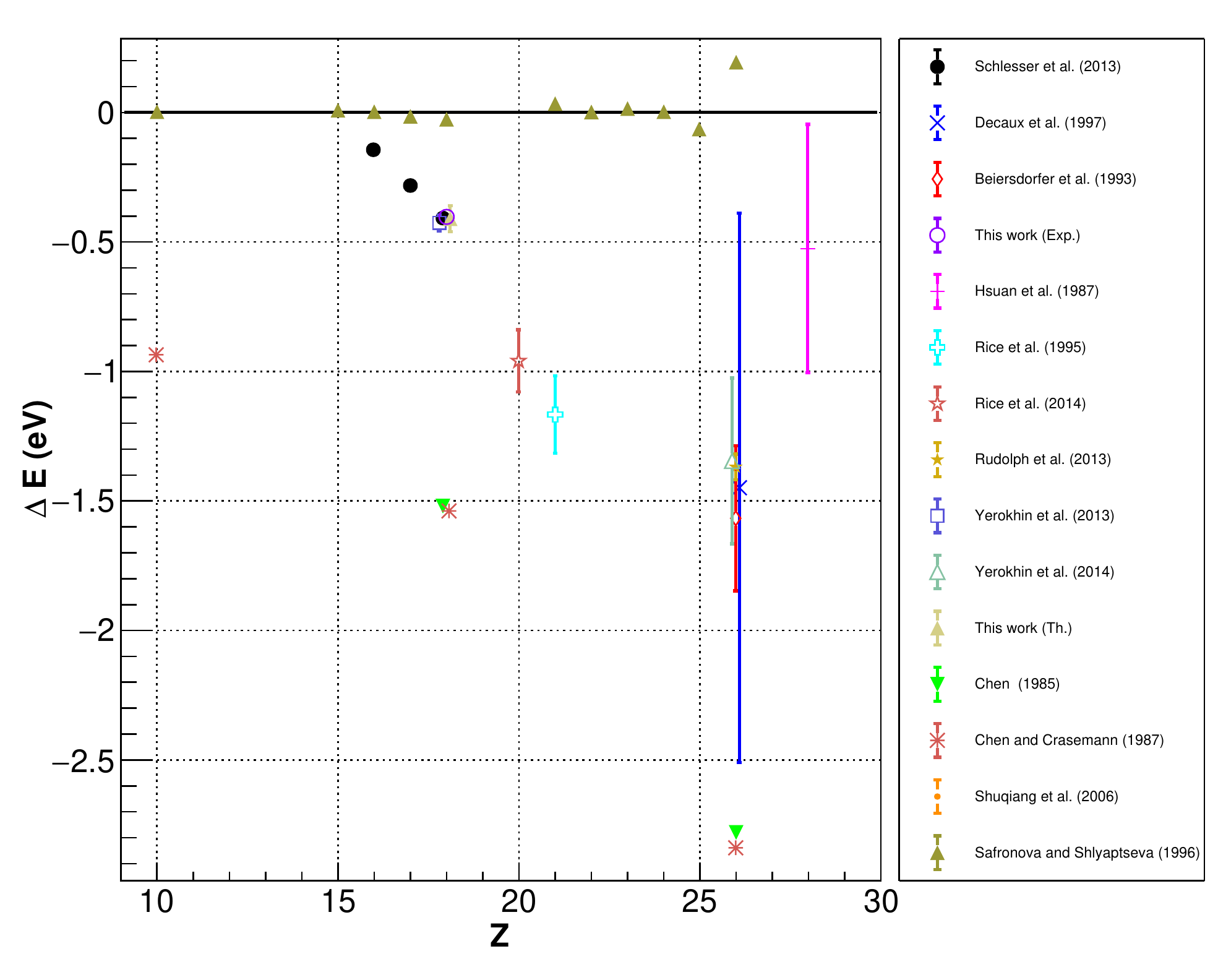}
\caption{
(Color online) Comparison between experimental and theoretical values for the  $1s2s^2 2p \,^1P_1 \rightarrow 1s^22s^2\,^1S_0$ transition energies, as a function of $Z$. 
All values are compared to the energies in Ref. \cite{sal1979}. 
The experimental results are from the following references: Schlesser   \etal (2013) \cite{sbcs2013}, Beiersdorfer  \etal (1993) \cite{bpjh1993}, Decaux   \etal (1997) \cite{dbkj1997}, Rudolph   \etal (2013) \cite{rbes2013}, Hsuan   \etal (1987) \cite{hbhv1987}, Rice   \etal (1995) \cite{rgtm1995}, Rice   \etal (2014) \cite{rrag2014}.
The theoretical results are from the following references: Yerokhin  \etal (2015) \cite{ysf2015}, Yerokhin \etal (2014) \cite{ysf2014}, Chen and Crasemann (1987) \cite{cac1987}, Chen (1985) \cite{che1985}, Shuqiang   \etal (2006) \cite{sfg2006}, Safronova and Shlyaptseva (1996) \cite{sas1996}.
\label{fig:be-z-comp}
}
\end{figure*}

\begin{table}							
\caption{							
Comparison between relative measurements from Ref. \cite{sbcs2013}, and the values deduced from this work and our previous measurement of the $1s2s\,^3S_1 \to 1s^2\, ^1S_0$ M1 transition \cite{asgl2012} for the $1s2p\,^1P_1 \to 1s^2\, ^1S_0$ and the $1s2s^2 2p \,^1P_1 \rightarrow 1s^2 2s^2\,^1S_0$ transition. All energies are given in eV, and  uncertainties are shown in parentheses. 
}
\label{tab:relat-ener} 							
\begin{ruledtabular}
\begin{tabular}{cdddd}
& 	\multicolumn{2}{c}{Experiment}	 & 	\multicolumn{2}{c}{Theory} \\				
Level	&	\multicolumn{1}{c}{This work, Ref. \cite{asgl2012} }		&	\multicolumn{1}{c}{Ref. \cite{sbcs2013}}		&	\multicolumn{1}{c}{Refs. \cite{ysf2015,asyp2005}}		&	\multicolumn{1}{c}{This work}	\\
\hline
\multicolumn{1}{c}{$1s 2p\,^1P_1$}	&	35.432	(10)	&	35.419	(11)	&	35.4337	(4)	&	35.434  \\
\multicolumn{1}{c}{$1s 2s^2 2p\,^1P_1$}	&	-12.383	(11)	&	-12.372	(3)	&	-12.399	(34)	&	-12.403	\\
%
\end{tabular}							
\end{ruledtabular}
\end{table}																			

%
\section{Conclusions}
\label{sec:conclusions}

In the present work, we report the reference-free measurement of two x-ray transition energies and widths in He-like ($1s2p \,^1P_1 \rightarrow1s^2\,^1S_0$) and Be-like ($1s2s^22p \,^1P_1 \rightarrow1s^22s^2\,^1S_0$) argon ions. 
The measurement of the $1s2s^22p \,^1P_1 \rightarrow1s^22s^2\,^1S_0$ transition energy is the first reference-free measurement for a transition of an ion with more than two-electrons. The measurements were made with a double-crystal spectrometer connected to an ECRIS. The data analysis was performed using a dedicated x-ray tracing simulation code that includes the physical characteristics and geometry of the detector.
The energy measurements agree within the error bars with the most accurate calculations and with other recent measurements. The measurement of the He-like transition is one of the 5 measurements with a relative accuracy below \num{~1E-5}. The measurement of the Be-like Ar transition is the first reference-free measurement on such a transition, and the only one with this level accuracy, except for measurements relative to nearby He-like transitions.

We have also performed MCDF calculations of the transition energies and widths, using both the MCDFGME code, with improved self-energy screening and the RCI flexible atomic code FAC and compared with all existing theoretical and experimental results available to us.
The MCDFGME theoretical results are in agreement with existing experimental results and with the most advanced calculations available.

We have analyzed the difference between all available $n=2 \to n=1$ experimental transition energies in He-like ions for $Z\geq 12$ and the theoretical results from Ref. \cite{asyp2005}. When taking into account the recent high-precision, reference-free measurements in heliumlike argon \cite{bbkc2007,asgl2012,kmmu2014} and the present result,  in He-like iron\cite{rbes2013},  and in He-like krypton\cite{esbr2015} from the Heidelberg and Paris groups, as well as the copper result ~\cite{bab2015} by the Livermore group, we have shown that there is no significant $Z$-dependent deviation between the most advanced theory and experiment. 

The method presented here will be extended to other charge-states like lithiumlike or boronlike ions, and nearby elements in the near future.
%
%
\begin{acknowledgments}
%
 This research was supported in part by the projects No. PEstOE/FIS/UI0303/2011, PTDC/FIS/117606/2010, and by the research centre grant No. UID/FIS/04559/2013 (LIBPhys), from FCT/MCTES/PIDDAC, Portugal. We acknowledge partial support from NIST (P.I.), from the PESSOA Huber Curien Program Number 38028UD and the PAUILF program 2017-C08. 
P.A., J.M., and M.G. acknowledge support from FCT, under Contracts No. SFRH/BPD/92329/2013, No. SFRH/BD/52332/2013, and No. SFRH/BPD/92455/2013 respectively.
%
Laboratoire Kastler Brossel (LKB) is ``Unit\'e Mixte de Recherche de Sorbonne Universit\'e, de ENS-PSL Research University, du Collège de France et du CNRS n$^{\circ}$ 8552''. 
P.I. is a member of the Allianz Program of the Helmholtz Association, contract n$^{\circ}$ EMMI HA-216 ``Extremes of Density and Temperature: Cosmic Matter in the Laboratory''.
The SIMPA ECRIS has been financed by grants from CNRS, MESR,  and  University Pierre and Marie Curie (now Sorbonne Universit\'e). The experiment has been supported by grants from BNM \emph{01 3 0002} and the ANR \emph{ANR-06-BLAN-0223}.
We wish to thank Jean-Paul Desclaux for his help improving the \emph{mcdfgme} code, and  Dr. Martino Trassinelli (INSP) for valuable discussions and his help during early stages of the experiment.

\end{acknowledgments}

%
%
%
\bibliography{ref2017}
\bibliographystyle{apsrev}

\end{document}